\documentclass{aa}  
\usepackage[varg]{txfonts}
\usepackage{subcaption}

\usepackage{threeparttable}
\usepackage{url}

\usepackage[breaklinks=true]{hyperref} 
    
\usepackage{color}
\usepackage[normalem]{ulem}

\begin{document}

    \title{Multiwavelength study of on-disk coronal-hole jets with IRIS and SDO observations}

   \author{M. Koletti
          \inst{1}, \inst{2}
          \and
          C. Gontikakis \inst{2}
        \and 
        S. Patsourakos \inst{3} 
        \and 
        K. Tsinganos \inst{1} 
          }

\institute{
    Department of Physics, National and Kapodistrian University of Athens, University Campus, Zografos, GR-157 84 Athens, Greece
    \and
    Research Center for Astronomy and Applied Mathematics, Academy of Athens, Soranou Efesiou Str., 4, 11527, Athens, Greece
    \and
    Physics Department, University of Ioannina, Ioannina 45110, Greece
}

   \date{Received; accepted 18 June 2024}

\abstract{Solar jets are an important field of study, as they may contribute to the mass and energy transfer from the lower to the upper atmosphere.} {We use the Interface Region
Imaging Spectrograph (IRIS) and Solar Dynamic Observatory (SDO) observations to study two small-scale jets (jet 1 and jet 2) originating in the same on-disk coronal hole observed in October 2013.}{We combine dopplergrams, intensity maps, and line width maps derived from IRIS Si IV 1393.755Å spectra along with images from the Atmospheric Imaging Assembly (AIA) on SDO
to describe the dynamics of the jets. Images from AIA, with the use of the emission measure loci technique and rectangular differential emission measure (DEM) distributions, provide estimations
of the plasma temperatures. We used the O \textsc{IV} 1399.77 \AA, 1401.16~\AA\ spectral lines from IRIS to derive electron densities.  
}{For jet 1, the SDO images show a small mini-filament 2 minutes before the jet eruption, while jet 2 originates at a pre-existing coronal bright point. The analysis of asymmetric spectral profiles of the Si
\textsc{IV} 1393.755~\AA\ and 1402.770~\AA\ lines reveals the existence of two spectral components. One of the components can be related to the background plasma emission originating outside the jet, while the secondary component represents higher-energy plasma flows associated with the jets. Both jets exhibit high densities of the order of 10$^{11}$ cm$^{-3}$ at their
base and 10$^{10}$ cm$^{-3}$ at the spire, respectively, as well as similar average nonthermal velocities of $\sim$ 50-60 km/s. However, the two jets show
differences in their length, duration, and plane-of-sky velocity. Finally, the DEM analysis reveals that both jets exhibit multithermal distributions.}{ This work presents a comprehensive description of the thermal parameters and the dynamic evolution of two jets.
The locations of the asymmetric profiles possibly indicate the areas of energy release triggering the jets.
} 


 \maketitle
%

\section{Introduction}

Jets are collimated plasma ejections that can serve as sources of mass and energy transfer to the upper atmosphere and the solar wind \citep{raoufi2016,cirtain_2007}. They occur everywhere on the Sun (in coronal holes (CHs), active regions, and the quiet Sun) and can be observed in a large range of temperatures (10$^4$ K such as chromospheric anemone jets \citep{shibata_2007}  - to 10$^6$ K) and wavelengths (optical, extreme-ultraviolet (EUV), and X-rays). 
In many cases, jets can originate in coronal bright points \citep{mou_2018}, which are small-scale loops observed in EUV and X-rays with typical sizes of 5\arcsec - 60\arcsec\ and maximum lifetimes of $\sim$ 20 hours, connecting areas of opposite magnetic polarity in the quiet Sun and coronal holes \citep{Madjarska2019}. 

Typically, jets consist of a straight beam-like spire, which is connected to a dome-like loop at its base \citep{Wyper_2019}. Usually, one of the footpoints of the loop is intensified during the jet ejection \citep{raoufi2016}. 
Generally, jets have apparent lengths of 10-100 Mm and reach speeds from 10 up to 1000 km/s, with an average value of 200 km/s \citep{Shimojo_1996}. They last from 1 minute up to several minutes \citep{cirtain_2007} and typical electron densities in the lower corona are of the order of 10$^9$ cm$^{-3}$ \citep{Doschek_2010}. Furthermore, coronal hole jets are faster compared to those observed in quite Sun regions \citep{Narang_2016}.

One proposed mechanism for the formation of solar jets is magnetic reconnection between open and closed magnetic field lines; however, the exact processes that trigger them are not yet fully understood.
Numerical models show that jets are triggered by magnetic reconnection occurring between the ambient field and emerging flux \citep{moreno_2008, Gontikakis_2009, archontis_2010}.
\cite{Muglach_2021} categorized 35 coronal jets based on the changes of the photospheric magnetic flux at their bases. These authors find that multiple cases can  exist: flux cancelation, complex flux changes, and no flux changes. Several studies suggest that jets are triggered by the eruption of mini-filaments, which lie on the polarity-inversion line between polarities
 of opposite sign (\cite{Panesar_2018},\cite{Li_2023}). However,  jets not associated with mini-filaments have also been observed \citep{Kumar_2019}. Another mechanism proposed for the formation of jets in the lower solar atmosphere, such as spicules, is the so-called whiplash effect, according to which magnetic tension is transported upwards, where it can drive flows of plasma and generate waves \citep{Sykora_2017}). 

Although detailed spectroscopic observations of jets are not as abundant as studies based on imaging observations, they can provide useful information about the properties of the emitting plasma. These properties include the line-of-sight Doppler velocity and the nonthermal width, with this latter being the result of several processes, such as waves and turbulent motions \citep{Chae_1998}. Jets exhibit blueshifted velocities in the transition region (TR) and the corona \citep{Scullion_2009}, indicating upward motions.
Moreover, asymmetries in TR and coronal spectral line profiles 
are interpreted as different plasma motions along the line of sight \citep{sykora_2011}. Such asymmetries have been observed in jets, revealing secondary components of blueshifted velocities and enhanced line broadenings \citep{Gorman2022}. 

Moreover, the ratio of two optically thin spectral lines originating from the same upper level can provide an estimate of the opacity of the emitting plasma \citep{Jordan1967}. 
Theoretically, the ratio of the resonance lines of Si \textsc{IV} formed at 1393.755~\AA\ and 1402.770~\AA\ is approximately equal to 2, assuming that the plasma is optically thin, while the plasma emission originates from free electrons and ions collisions \citep{DelZanna2018}. Values smaller than 2 are an indication of opacity effects \citep{Buchlin_Vial_2009,Dere_1993} and are attributed to increased density along the line of sight \citep{Tripathi_2020}. Ratios above 2 are thought to be a result of resonant scattering  or can be due to the geometry and orientation of the structure toward the line of sight \citep{Keenan_2014}.  

Resonant scattering can dominate plasma thermal emission of a given structure, when a bright nearby source illuminates it, such as a microflare \citep{Gontikakis_2013}. Resonant scattering can be a non-negligible emission process when there is a combination of a relative high incident intensity that illuminates the structure, while the thermal emission is depressed relative to a typical thermal emission value for these spectral lines \citep{Gontikakis2018}. A low thermal emission is the result of reduced electron density or of an electron plasma temperature that is not equal to the temperature corresponding to the maximum Si~\textsc{IV} ion population (which, for this ion, corresponds to 10$^{4.9}$~K in ionization equilibrium \citet{Doschek_1997}).

Here, we present a detailed study of two small-scale jets observed within an on-disk coronal hole by the Interface Region Imaging Spectrograph (IRIS) and Solar Dynamic Observatory (SDO). IRIS data have been widely analyzed for the study of solar jets, as thoroughly discussed in \cite{Schmieder_2022}. We note that most studies focus on active regions. \cite{Panesar_2022} analyzed Mg \textsc{II} spectra and find clear evidence of flux cancelation associated with the eruption of a mini-filament triggering a jet.

This paper is organized in the following way: In \autoref{observation}, we present the observational data. In \autoref{data analysis and results} we describe the data analysis and the main results, while in \autoref{discussion} we summarize our main conclusions.

\section{Observations} \label{observation}

We studied two small-scale jets occurring within an on-disk coronal hole (CH) region and observed in the chromosphere, TR, and the corona on October 9, 2013. 
We analyzed data obtained by IRIS \citep{DePontieu_2014}, which provides very high spatial (0.35\arcsec ) and spectral (40 m\AA) resolution data in the far-ultraviolet (FUV 1332\AA\ - 1407 \AA). We used slitjaw images (SJ) at 1330~\AA\ and 1400~\AA dominated by the  C \textsc{II} 1334.532/1335.707~\AA\ and Si \textsc{IV} 1393.755/1402.770~\AA\ emissions, respectively, as well as a very dense raster scan of 3.5 hours in duration and 400 slit positions.
The observation is centered on $[$X,Y$]$=$[$435.2,302.6$]$\arcsec\ from the center of the solar disk with a field of view of $[$140.5,181.8$]$\arcsec\ and lasts from 16:56 UT to 20:26 UT. 
We analyzed level-2 data, which have been corrected for various effects (e.g., flat-field, dark current) and have been wavelength calibrated with respect to the chromospheric O I 1955.6 \AA\ line.

\begin{figure}
    \centering
\resizebox{\hsize}{!}{\includegraphics{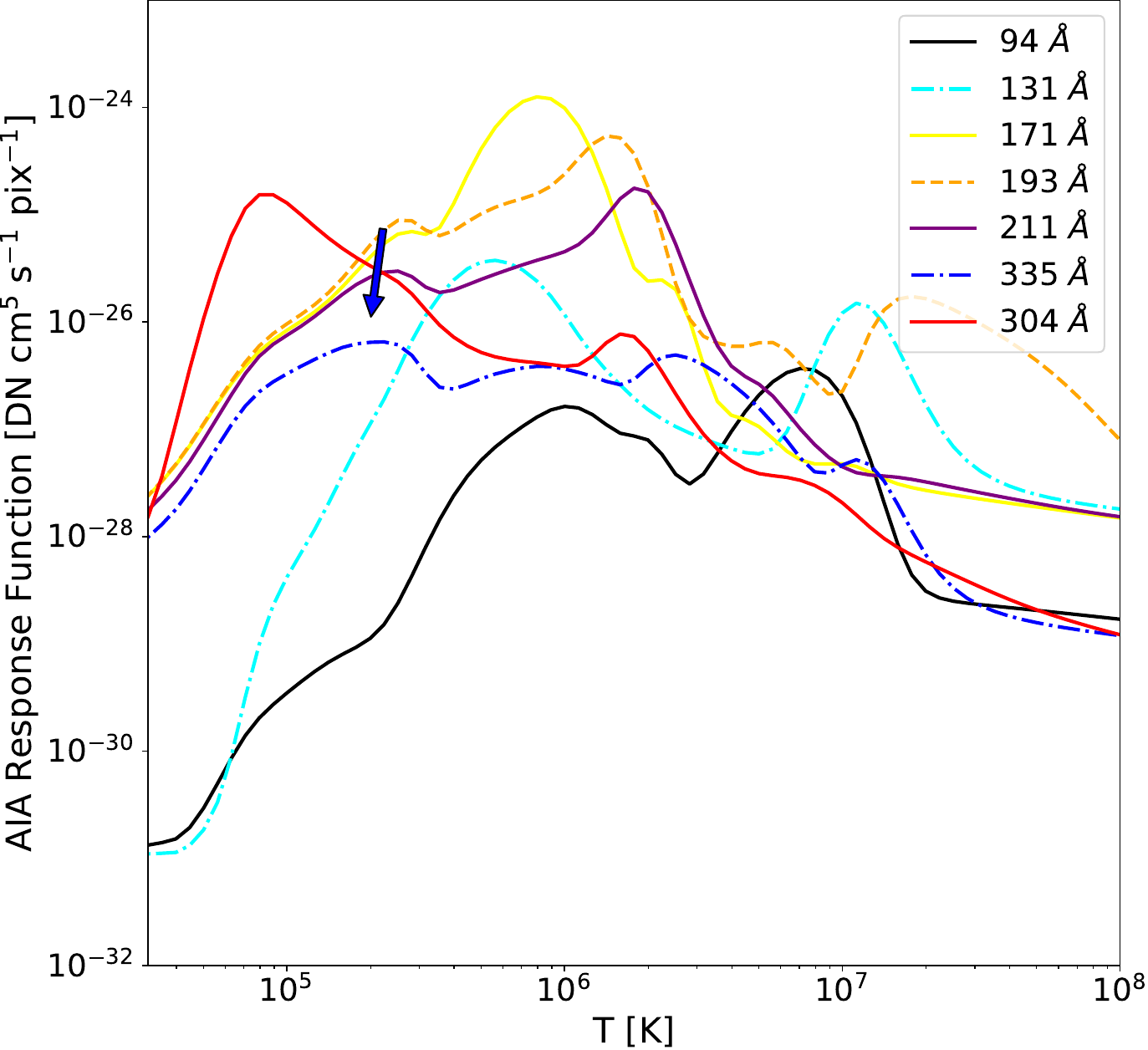}}
        \caption{Response functions of the AIA channels calculated with CHIANTI v.10.0.1. The blue arrow shows the peak response of the 335~\AA\ channel, which corresponds to a lower temperature than that associated with the primary ion mentioned in the text.}
    \label{aia_response_function}
\end{figure}

Moreover, we used EUV data (as recorded by the Atmospheric Imaging Assembly, AIA,  \cite{Lemen2012}) and line-of-sight magnetograms (LOS, as recorded by the Helioseismic Magnetic Imager, HMI, \cite{Scherrer_2012}). Both AIA and HMI are instruments on board the Solar Dynamic Observatory (SDO).
The AIA data have a cadence of  12 seconds
 and a pixel size of  0.6\arcsec\ , while the HMI magnetograms are of the same pixel size and have a  cadence of 45 seconds, with an upper bound of the random noise equal to 10 Gauss \citep{liu_2012}. 
From the AIA instrument, we select the following channels: 304~\AA\ (primary ion: He~\textsc{II}, characteristic temperature: logT=4.7 K), 131~\AA\ (Fe~\textsc{VIII}, logT=5.6 K), 171~\AA\ (Fe~\textsc{IX}, logT=5.8 K), 193~\AA\ (Fe~\textsc{XII},  logT=6.1 K),  211~\AA\ (Fe~\textsc{XIV}, logT=6.3K), 335~\AA\ (Fe~\textsc{XVI}, logT=6.4K), and 94~\AA\ (Fe~\textsc{XVIII}, logT=6.8K). It should be emphasized that multiple ions contribute to the total emission of each channel. For example, ions forming in lower temperatures of around 10$^{5.3}$ K contribute to the 335~\AA\ channel of temperatures, as shown in \autoref{aia_response_function}.

\begin{figure}
    \centering\resizebox{\hsize}{!}{
\includegraphics{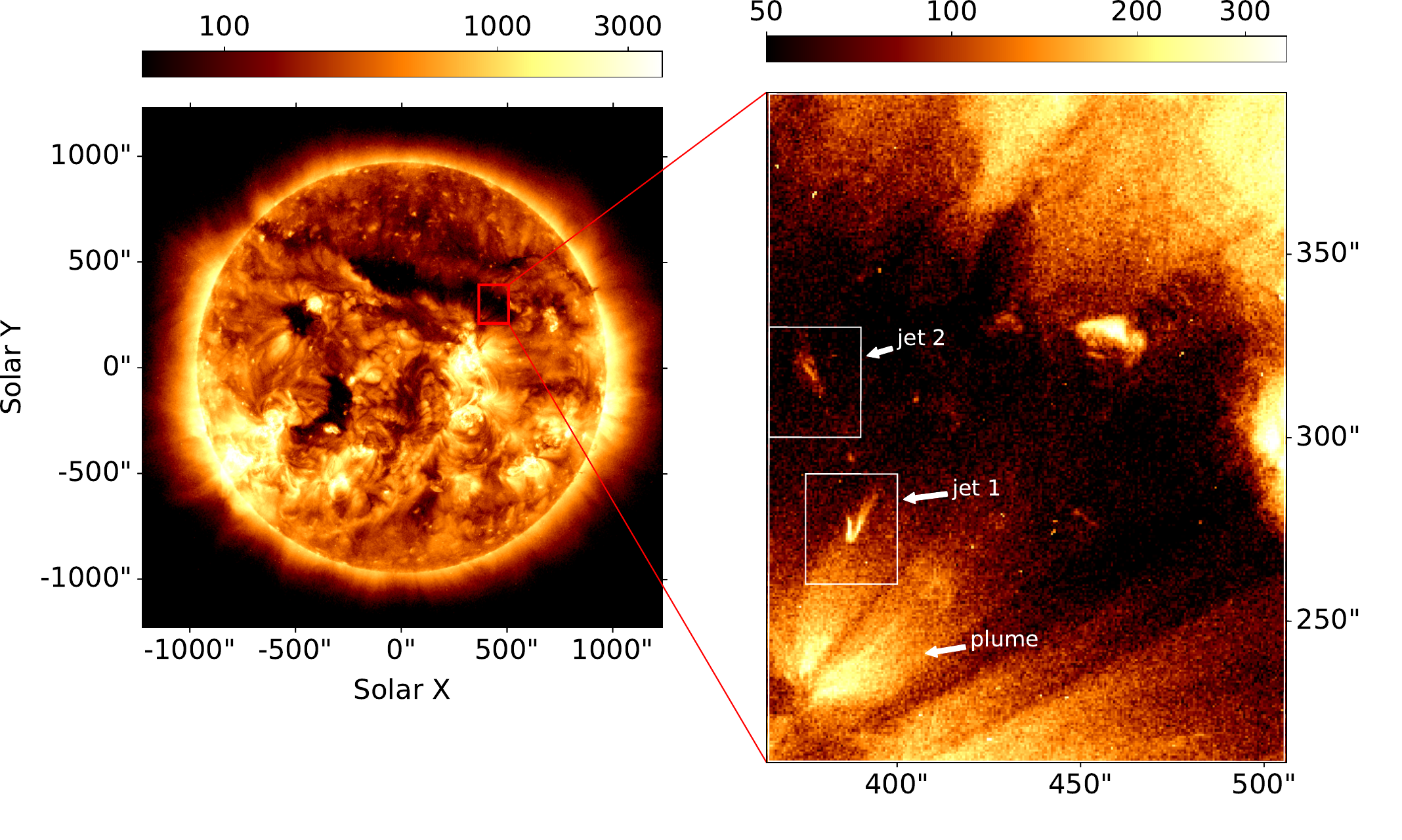}}
        \caption{Left panel: the position of the CH under study on October 9, 2013 at 17:27 UT. Right panel: the two jets under study (white boxes). On the bottom left corner the white arrow shows the base of the plume, which is the foreground of jet 1. The images are obtained by AIA at 193~\AA\ in DN px$^{-1}$ s$^{-1}$.}
    \label{aia_cut_out}
\end{figure}

\autoref{aia_cut_out} shows the CH on the solar disk (left panel), as well as the two areas where the jets under study occur, which are  enclosed with white boxes (right panel). In the bottom left corner, the white arrow shows the root of the plume, which serves as foreground for jet 1. 
The image was obtained by AIA at 193~\AA\ at 17:27~UT.

\section{Data analysis and results} \label{data analysis and results}

\subsection{Coalignment and data preparation}

The AIA images are co-aligned with the IRIS observation by visually comparing the AIA 1700~\AA\ and IRIS 1330~\AA\ slitjaw images of similar time stamps. The HMI magnetograms are co-aligned with the 1700~\AA\ AIA image in the same way. The error can be estimated to 2-3 pixels (1-2\arcsec ). 
In order to enhance the signal of the HMI magnetograms and reveal weak magnetic structures, we took the average of three consecutive magnetograms.

For the treatment of the AIA Level-1 to Level-1.5 images, we used version 0.6.3 of the aiapy open source software package \citep{Barnes2020}, with which we apply the following: image deconvolution on the AIA images with the instrument point spread function using the Richardson-Lucy deconvolution algorithm; update of the image pointing; scaling of the images to 0.6~\arcsec\ per pixel; correction of the image rotation relative to the solar north; and finally normalization of the exposure time to have intensity
units of DN per pixel per second, where DN denotes data number.

\subsection{Description of the jets} \label{description}

\subsubsection{Jet 1}

\begin{figure*}
    \centering
     \includegraphics[width=17cm]{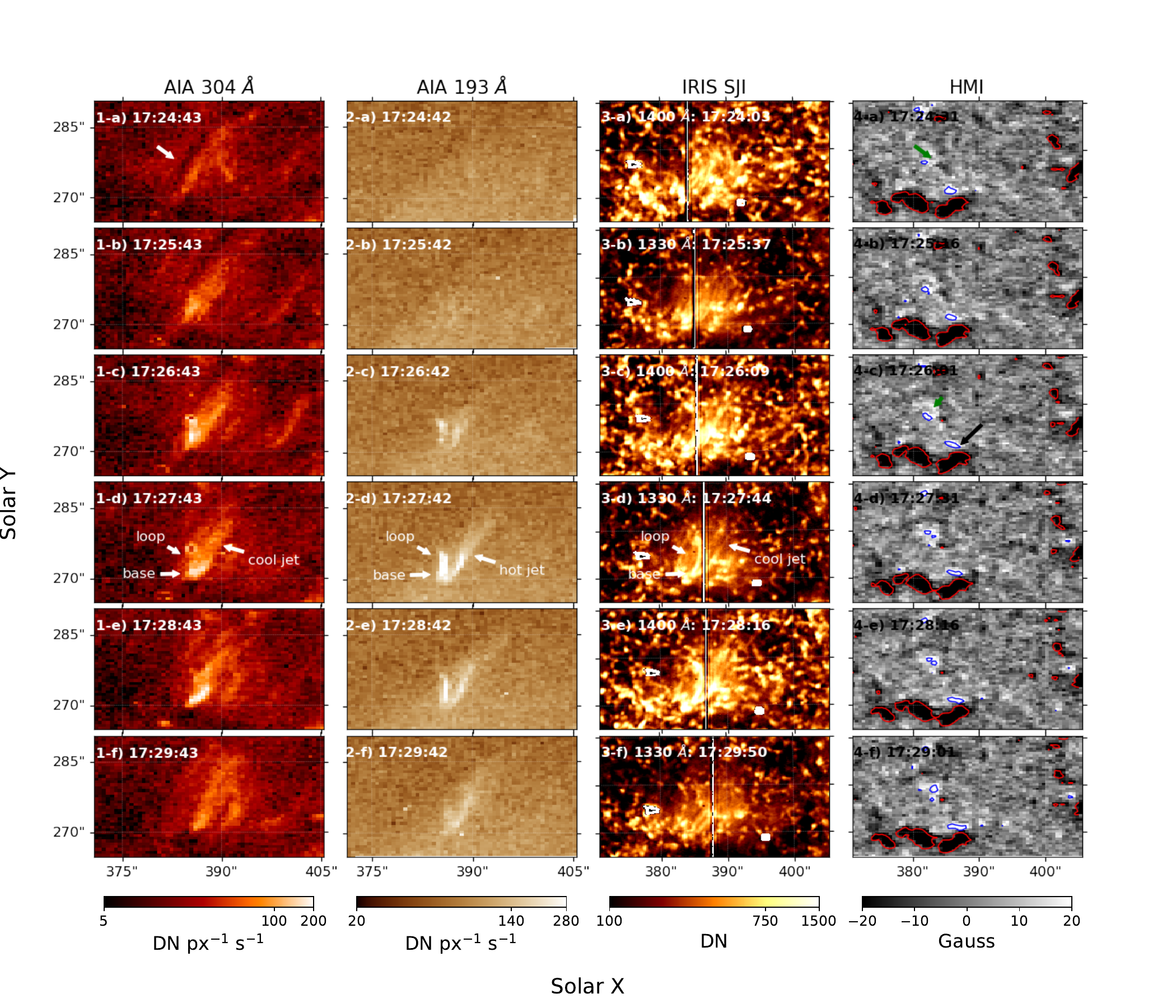}
    \caption{Temporal evolution of Jet 1. Columns 1 to 4 show AIA 304~\AA\ and 193~\AA, SJ 1330~\AA\ and\   1400~\AA,\ and HMI magnetogram images, respectively. Time evolves from 17:24 UT (first row) to 17:29 UT (last row). The arrows in panels 1-d and 3-d indicate the location of the loop and the jet under study. In panels 4-a to 4-f, the red and blue contours indicate regions with strengths of -20G and 20G, respectively. The white arrow in panel  1-a shows the position of the mini-filament (see text for more details) and the corresponding green arrow in panel 4-a indicates the underlying polarity inversion line.}
    \label{jet 1}
\end{figure*}

Jet 1 is observed in the 171~\AA, 131~\AA, 193~\AA, 211~\AA, 304~\AA, and 335~\AA\ AIA channels, as well as in the 1330~\AA\ and 1400~\AA\ IRIS slitjaw images. \autoref{jet 1} shows the evolution of the jet at the 304~\AA\ and 193~\AA\ AIA channels (panels 1-a to 2-f), at the 1330~\AA\ and 1400~\AA\ IRIS SJ images (panels 3-a to 3-f) and the HMI magnetogram (panels 4-a to 4-f), where the red and blue contours correspond to -20~G and 20~G, respectively.
The jet also consists of two counterparts of different temperature, a cool counterpart observed in 304~\AA\ and the SJ images (panels 1-a to 1-f and 3-a to 3-f) of \autoref{jet 1}, respectively), as well as one of higher temperature in the corona (panels 2-a to 2-f) of \autoref{jet 1}). 

A loop-like structure appears at the same time as the jet and is also observed in all AIA channels except 94 \AA,\, and in the 1330 \AA\ and 1400 \AA\ IRIS SJ images. The spire and the loop appear to be connected at the same bright point at their base. 
Four structures are indicated with white arrows in panels 1-d, 2-d, and 3-d of \autoref{jet 1}, which correspond to the loop, the jet base, the cool jet in 304~\AA,\ and the hot jet in 193~\AA.  
\autoref{aia_cut_out} shows a plume rooted at the edges of the coronal hole, which is projected over jet 1. The plume is not physically related to the jet but its emission is a foreground over the jet emission.

The cool component of the jet first appears in the 304~\AA\ images approximately 2 minutes prior to the coronal images, and therefore has a lifetime of approximately 6 minutes, from 17:24~UT to 17:30~UT. By visually examining both the chromospheric (304~\AA) and coronal (193~\AA) images, we find plane-of-sky lengths of 10.5 Mm and 4.5 Mm for the jet and the loop, respectively. The width of the jet remains constant throughout its lifetime and approximately equal to 2 Mm when observed in all wavelengths except 304~\AA, in which it appears to increase in width and change shape. We also estimated the plane-of-sky velocity from the 193~\AA\ AIA images, finding a value of $\approx$100~km/s.

The HMI magnetograms (\autoref{jet 1} panels 4-a to 4-f) reveal that the underlying photospheric magnetic field is mostly unipolar. However, there is a patch of opposite polarity with a strength of $\approx$ 50 G close to the base of the jet, at approximately (X,Y)=(390\arcsec ,272\arcsec ), which appears to move toward the negative polarity (shown with the black arrow in panel 5-c). Moreover, a positive polarity appears at (X,Y)=(387\arcsec ,287\arcsec ), which could be connected with the loop (green arrow at panel 4-c).

Interestingly, a dark feature of  $\sim$15\arcsec\  in length appears 2 minutes prior to the jet eruption in 304~\AA\ and 193~\AA\  (panels 1-a and 2-a \autoref{jet 1}, indicated with a white arrow) and disappears thereafter. This feature is absent from  the 1400~\AA\ image, suggesting that the feature is a cool mini-filament containing neutral-hydrogen-absorbing material.
The mini-filament exhibits sizable proper motions in the AIA images between 17:23:55~UT and 17:24:31~UT at the beginning of the jet~1 launch, referring to a mini-filament eruption, similarly to \cite{sterling_2015}. However, we cannot distinguish whether
it is ejected from the solar surface or is confined inside a close magnetic structure \citep{Sterling_2022} because of the low spatial resolution and the fact that it is observed on the disk.
 The corresponding magnetic field configuration (panel 4-a \autoref{jet 1}) shows that the mini-filament lies above two patches of opposite polarity  with total strengths of approximately 20 G and -20 G (shown on \autoref{jet 1} 4-a panel with a green arrow),  forming a weak polarity inversion line. However, the magnetic field magnitude of these patches is very weak.

 \subsubsection{Jet 2}

\begin{figure*}
 \includegraphics[width=17cm]{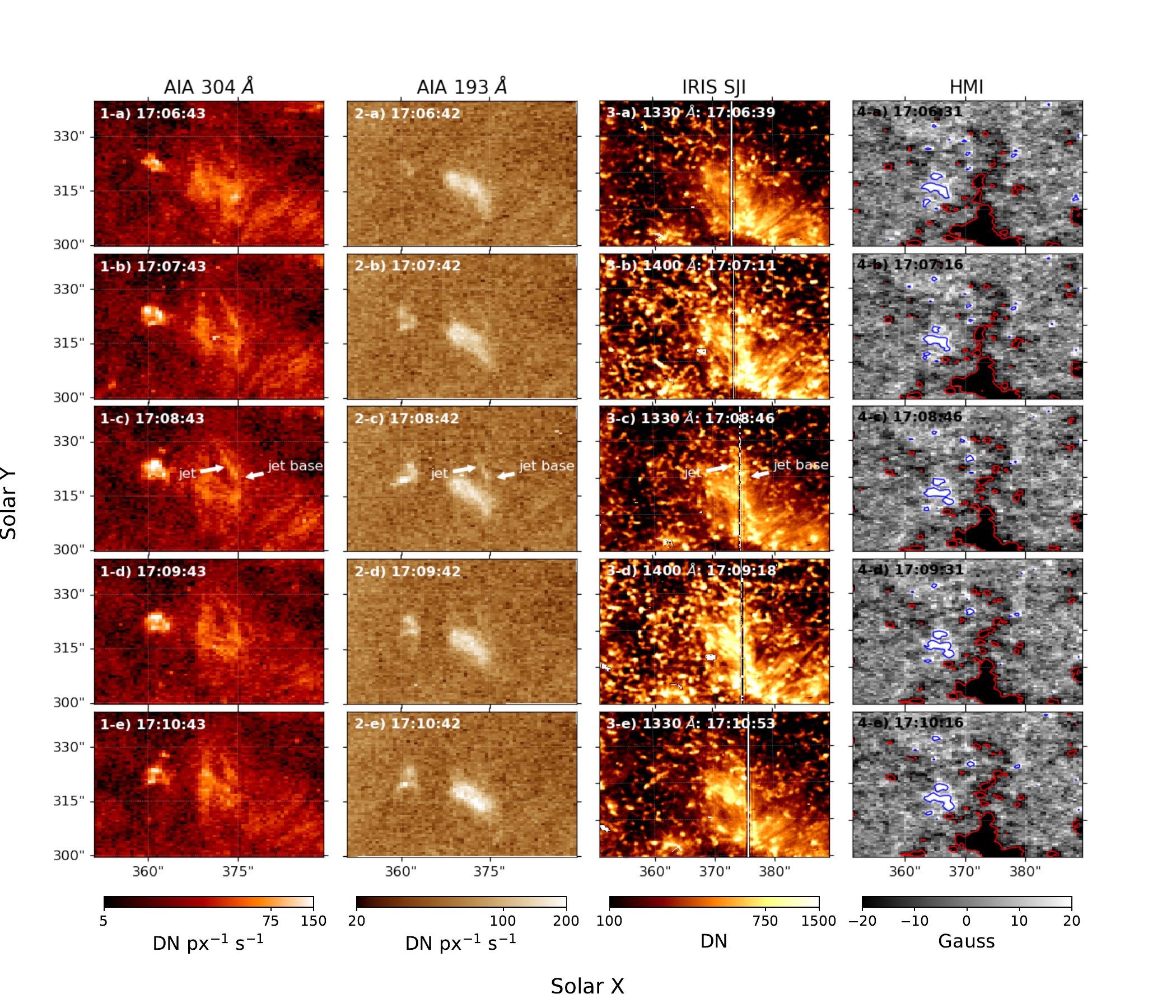}  \caption{Temporal evolution of Jet 2. Columns 1 to 4 show AIA 304~\AA\ and 193~\AA, SJ 1330~\AA\ and 1400~\AA, and HMI magnetogram images, respectively. Time evolves from 17:06 UT (first row) to 17:10 UT (last row). The white arrow in panels 1-c, 2-c, and 3-c shows the jet eruption. The red and blue contours indicate regions with strengths of -20G and 20G, respectively.}
    \label{jet 2}
\end{figure*}

Jet 2 occurs at 17:08 UT, and originates from a pre-existing coronal bright point and lasts about 3-4 minutes. \autoref{jet 2} shows the evolution of the coronal bright point and the jet at the 304~\AA\ and 193~\AA\ AIA channels (panels 1-a to 2-e), at the 1330~\AA\ and 1400~\AA\ IRIS SJ images (panels 3-a to 3-e) and the HMI magnetogram (panels 4-a to 4-e), where the red and blue contours correspond to -20~G and 20~G, respectively. The location of the jet and its base are indicated with white arrows in panels 1-c, 2-c, and 3-c. The jet spire has a length of 9 Mm and a width of 1.5 Mm and the apparent plane-of-sky velocity derived from the 193~\AA\ AIA images is $\approx$80km/s.  We note that we do not observe any evolution in the magnetic field that could be associated with the jet eruption, as any variation of its strength is within the noise levels.

\subsection{Fitting of spectral profiles} \label{fitting}

We analyzed the IRIS spectral lines Si \textsc{IV} 1393.755~\AA\ and Si \textsc{IV} 1402.770~\AA\  by applying Gaussian fits. Profiles affected by cosmic rays were treated with a simple despiking algorithm, which linearly interpolates the corrupted spectral points with their nearest neighbors. 
We find asymmetric profiles concentrated along the loop and at the base of jet 1 (see \autoref{rad jet 1}), and at the base of jet 2.  Interestingly, we do not find asymmetric profiles along the spires of either jet, but only at their bases and the loop associated with jet 1. It should be noted that the IRIS slit captures the spire of jet 1 only at its descending phase. 

For both jets, the asymmetric profiles are fitted with a bimodal Gaussian curve (\autoref{BG fit}), from which we derive two components: a more intense, narrower and less Doppler-shifted "core" component and a wider and blueshifted or redshifted "tail" component.  In all cases, the tail component is characterized by a larger spectral width compared to the core component. We distinguish the asymmetric profiles based on visual inspection, ensuring also that, in most cases, the $\chi^2$ of the bimodal Gaussian fit is smaller than the $\chi^2$ of the single Gaussian fit. 
Taking into account that the two spectral lines Si~\textsc{IV} 1393.755~\AA\ and Si~\textsc{IV} 1402.770~\AA\ are emitted by the same plasma volume, we apply constraints to the fitting process, so that the spectral width and Doppler shift are the same for the core components of the two lines. The same applies for the tail components.

\begin{equation} \label{BG fit}
   I=I_{cont}+I_{CC}\cdot e^{-\frac{(\lambda-\lambda_{CC})^{2}}{\sigma_{CC}^2}} +I_{TC}\cdot e^{-\frac{(\lambda-\lambda_{TC})^{2}}{\sigma_{TC}^2}}
,\end{equation}

where $I_{CC}$, $\lambda_{CC}$, and $\sigma_{CC}$ are the peak of the line, the centroid position of the profile relative to the rest wavelength, and the standard deviation of the core component, respectively. Similarly,  $I_{TC}$, $\lambda_{TC}$, and $\sigma_{TC}$ correspond to the tail component. $I_{cont}$ is the intensity of the constant continuum. 
The remaining profiles were fitted with a single Gaussian (SG) fit:

\begin{equation}
I=I_{cont}+I_{0}\cdot e^{-\frac{(\lambda-\lambda_0)^{2}}{\sigma^2}}
.\end{equation}

\begin{figure}
\centering
 \resizebox{\hsize}{!}{
\includegraphics{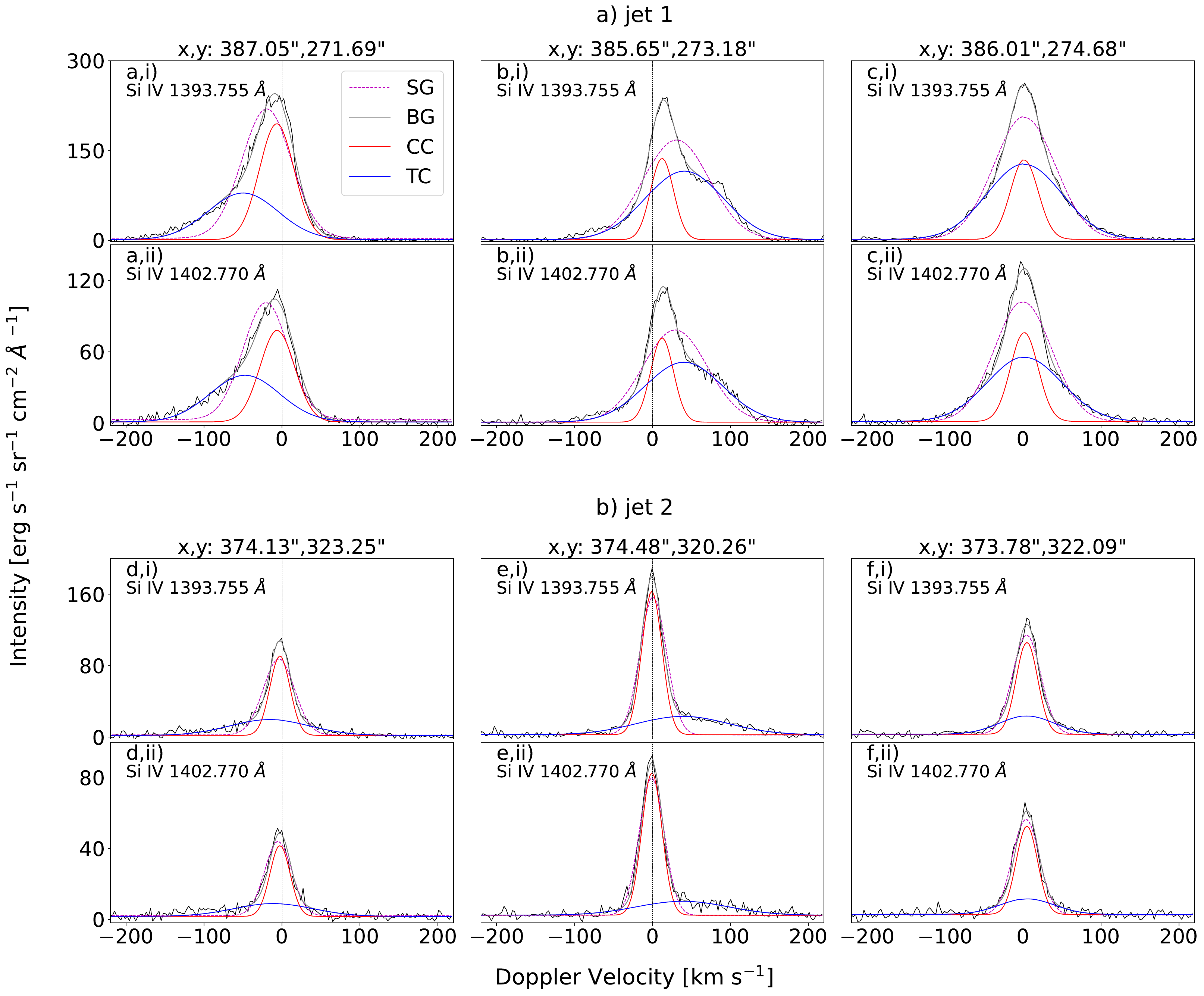}}
\caption{Characteristic Si \textsc{IV} 1393.755~\AA\ and 1402.770~\AA\ profiles of jet 1 (top subfigure) and jet 2 (bottom subfigure). The x-axis is the Doppler velocity in km/s and the y-axis is the intensity in erg s$^{-1}$ sr$^{-1}$ cm$^{-2}$ \AA$^{-1}$. In each panel, the red dashed line is the single Gaussian curve, the gray solid line is the bimodal Gaussian curve, and the red and blue solid lines are the core component and tail component curves, respectively. For each jet subfigure, the first row shows Si \textsc{IV} 1393.755~\AA\ profiles and the second row shows the respective Si~\textsc{IV} 1402.770~\AA\ profiles. The profile spatial position in arcseconds is indicated at the top of the Si \textsc{IV} 1393.755~\AA\ 
panels.} 
\label{profile_examples}
\end{figure}

All fitting procedures were performed with the Python lmfit.Model package \citep{lmfit}. 
Characteristic bimodal Gaussian profiles are shown in \autoref{profile_examples}. The top set of 6 panels correspond to jet~1, while the bottom set corresponds to jet~2. The x-axis is the Doppler velocity in km/s and the y-axis is the intensity in erg s$^{-1}$ sr$^{-1}$ cm$^{-2}$ \AA$^{-1}$. In each panel, the red dashed line is the single Gaussian curve, the gray solid line is the bimodal Gaussian curve, and the red and blue solid lines are the core component and tail component curves, respectively. It can be clearly seen that these profiles cannot be characterized by a single Gaussian curve. Moreover, both the core component and tail component for jet 1 are more intense than for jet 2. For both regions, the core component features are significantly less Doppler shifted, while the tail component features exhibit redshifts and blueshifts, but also in some cases no Doppler shift (panels c(i-ii), and f(i-ii)). These points are discussed in sections.

\autoref{raster} shows the maps of the total linear intensity (\autoref{rad jet 1} and \autoref{rad jet 2}) in erg s$^{-1}$ sr$^{-1}$ cm$^{-2}$, the Doppler velocity (\autoref{doppler jet 1} and \autoref{doppler jet 2}) in km/s, the nonthermal velocity (\autoref{nonthermal jet 1} and \autoref{nonthermal jet 2}) in km/s, and the Si~\textsc{IV} 1393.75~\AA\ over Si~\textsc{IV} 1402.77~\AA\ intensity ratios (\autoref{ratio jet 1} and \autoref{ratio jet 2}) for jet 1 and jet 2. The maps are composed of the single Gaussian parameters, except where the bimodal Gaussian parameters are indicated with the green x-marks in \autoref{rad jet 1} and \autoref{rad jet 2}. In the left panel of each subfigure, the values of the green x-marked pixels are those of the core component of the bimodal Gaussian fit, while the right panel shows those of the tail component. The calculations and analysis regarding each physical parameter mentioned above are discussed in the following subsections. 

\subsection{Intensity} \label{intensity}

In order to check if the core components from the bimodal profiles are emitted by a background plasma or from different co-spatial structures belonging to the jet and the loop,  we defined mean background profiles. 
We define the background as the regions enclosed with the gray boxes in \autoref{rad jet 1} and \autoref{rad jet 2}. The average of the background intensity distribution is $\sim$ 800 erg s$^{-1}$ sr$^{-1}$ cm$^{-2}$ for both jets.

We observe that the core-component intensities are comparable with the background for both jets in most cases, meaning that they represent the background emission. On the other hand, the average total intensity of the tail components is higher than that of the core components, indicating that the tail components are emitted by plasma associated with the jet eruptions.

The average tail component intensity of jet 1 is almost two times higher than that of jet 2; however the maximum values are of the same order for the two regions. Higher intensities are observed along the loop associated with jet 1 as well as at the coronal bright point.

\subsection{Velocity and nonthermal velocity} \label{dv ntv}

We calculate the line-of-sight Doppler velocity and the nonthermal velocity from the single Gaussian fit of the Si \textsc{IV} 1393.755~\AA\ spectral line, as well as for the two components derived from the bimodal Gaussian fit.   

The nonthermal velocity v$_{nth}$ is calculated as 

\begin{equation}
    v_{nth}= \frac{c}{\lambda} \cdot \frac{w_{nth}}{2 \sqrt{ln2}}
\end{equation}

where w$_{nth}$ is the nonthermal width,
$\lambda$ the observed wavelength, and $c$ is the speed of light. More details on the specific calculations can be found in the IRIS documentation \footnote{\url{https://iris.lmsal.com/itn38/analysis_lines_iris.html\#line-fitting}}. We use the formation temperature of the Si \textsc{IV} ion, which is equal to 10$^{4.9}$ K \citep{Doschek_1997}, equating to a thermal velocity of 4 km/s. 

For jet 1, the strongly blueshifted area in \autoref{doppler jet 1} for both tail component and core component profiles outlines the base of the jet forming a box with
lower left coordinates (X,Y)=(386\arcsec, 270\arcsec) to upper right
coordinates=(388\arcsec, 272\arcsec). In this area, the Doppler velocity of the core component and tail component  are both blueshifted and reach values of 20 and 80 km/s towards the observer, respectively.
The fact that both spectral components are blueshifted (on the contrary to the fact that the single Gaussian
background profiles defined in \autoref{intensity} are redshifted) indicates that the two spectral components of the jet 1 spire may belong to different strands forming jet 1.

The redshifted profiles arranged along the slit, from (386\arcsec, 270\arcsec) to  (386\arcsec, 278\arcsec), are the footpoints of the loop, where the Doppler velocity of the tail component is up to 40km/s away from the observer, while the Doppler velocity of the core component is almost equal to or lower than the Doppler velocity measured in profiles located outside the jet and described by a single Gaussian fit that we define as background in \autoref{intensity}. This
indicates that the loop is sampled only by the tail component profiles. Indicatively, the average Doppler velocity of both the core component and single Gaussian profiles is approximately equal to 5km/s away from the observer. Moreover, we measure significantly higher nonthermal velocity for the tail component of the bimodal Gaussian profiles that are located along the spire and the base of the jet, up to 130 km/s and 90 km/s, respectively (\autoref{nonthermal jet 1}).

Regarding jet 2, the faintly blueshifted region at (375\arcsec, 325\arcsec) in \autoref{doppler jet 2} (indicated with the black arrow) corresponds to the jet spire. The
individual profiles forming the spire are described by single Gaussian
and have a Doppler velocity of approximately 5 km/s toward the
observer; except for a small number of profiles that are described by a bimodal Gaussian, where the tail component reaches blueshifted Doppler velocities of $\sim$ 30 km/s. This area can be interpreted as the start of the jet.  
Below this region, at the base of the jet at (375\arcsec, 320\arcsec), there are a significant number of bimodal Gaussian profiles that exhibit high tail-component redshifted Doppler velocities of up to 40 km/s, while the core component Doppler velocity is almost comparable to the background and is equal to 2km/s. The nonthermal velocity of the tail components reaches 110 km/s at both the base and the start of the jet spire (\autoref{nonthermal jet 2}).

\subsection{Opacity} \label{opacity}
\begin{figure*}[hbt!]
    \centering
    
    \begin{subfigure}[c]{0.25\linewidth}
        \centering
        \caption{Jet 1 Total Intensity}
        \label{rad jet 1}
        \includegraphics[width=\linewidth]{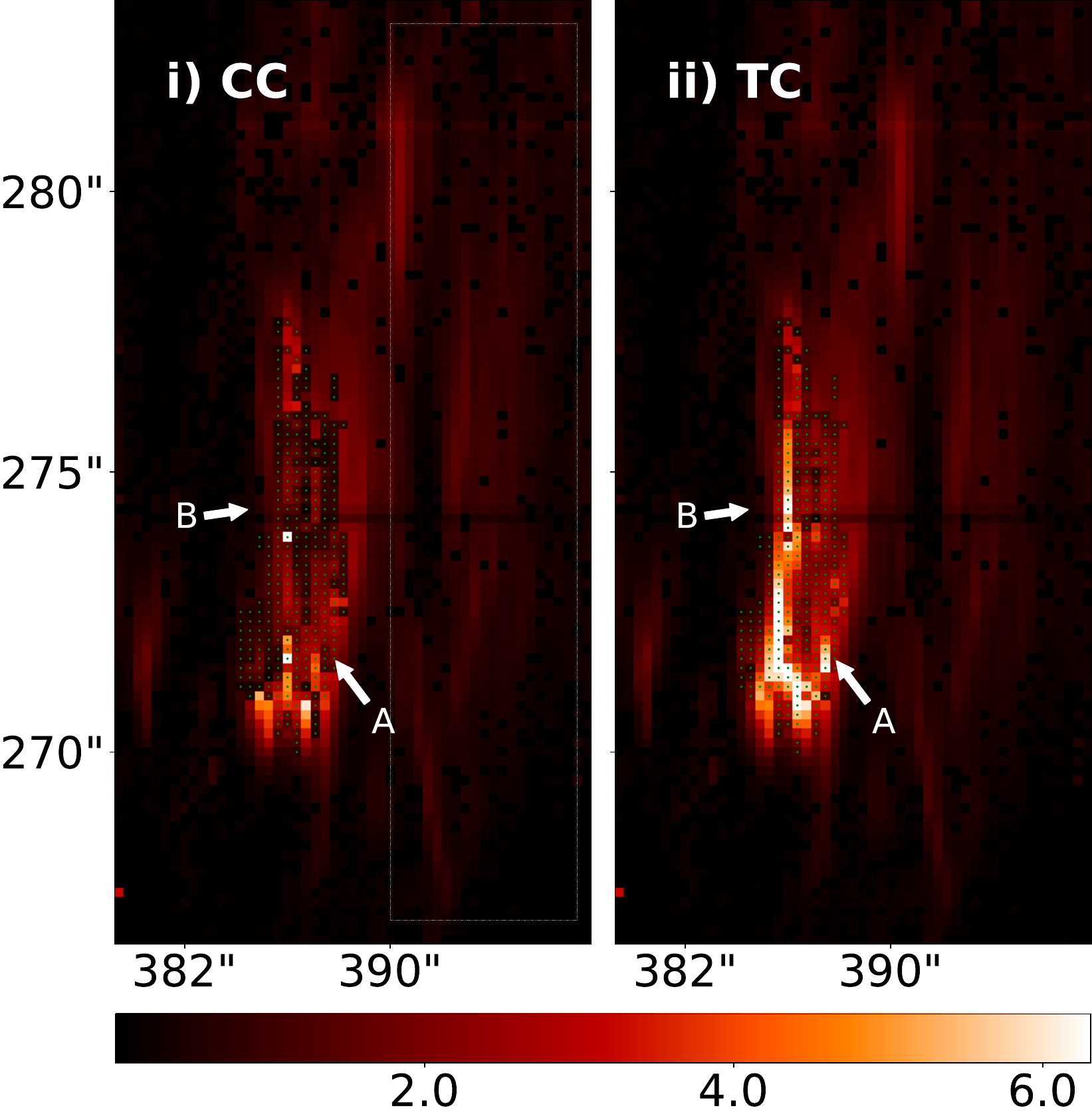}
    \end{subfigure}
    \begin{subfigure}[c]{0.225\linewidth}
        \centering
        \caption{Jet 1 Doppler Velocity}
        \label{doppler jet 1}
         \includegraphics[width=\linewidth]{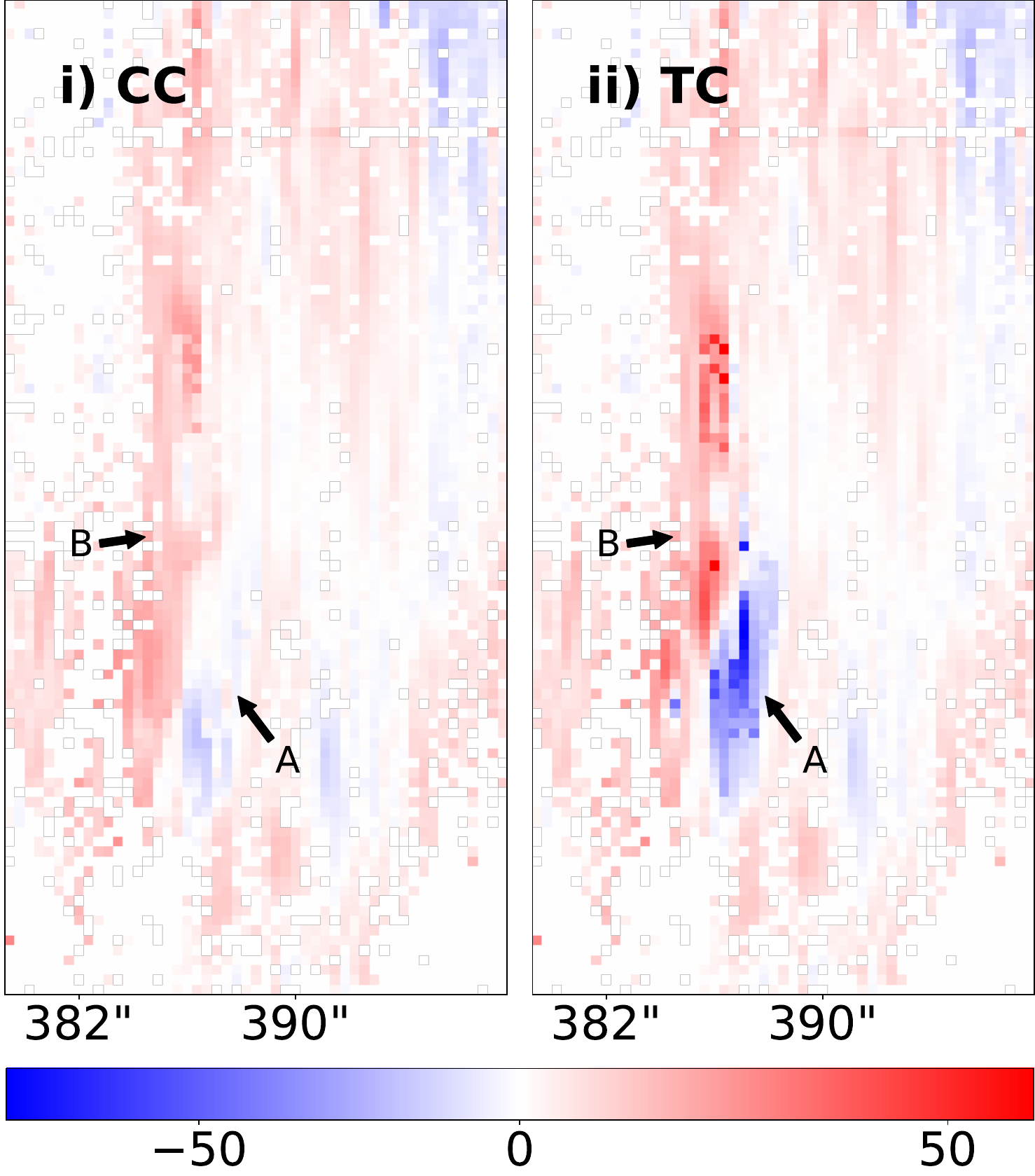}
    \end{subfigure}
    \begin{subfigure}[c]{0.225\linewidth}
        \centering
        \caption{Jet 1 Nonthermal Velocity}
        \label{nonthermal jet 1}
         \includegraphics[width=\linewidth]{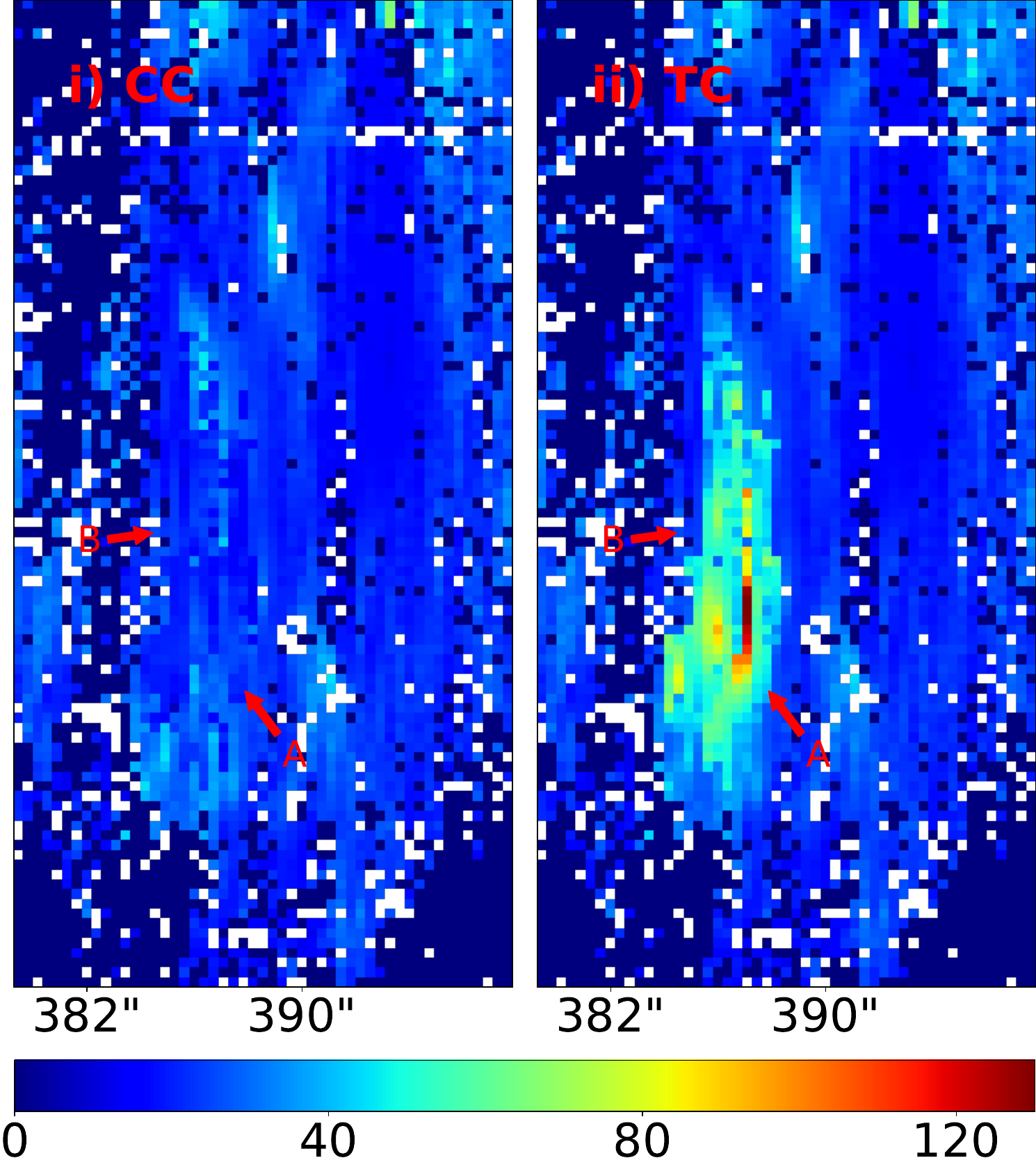}
    \end{subfigure}
    \begin{subfigure}[c]{0.23\linewidth}
        \centering
        \caption{Jet 1 Ratio}
        \label{ratio jet 1}
                \includegraphics[width=\linewidth]{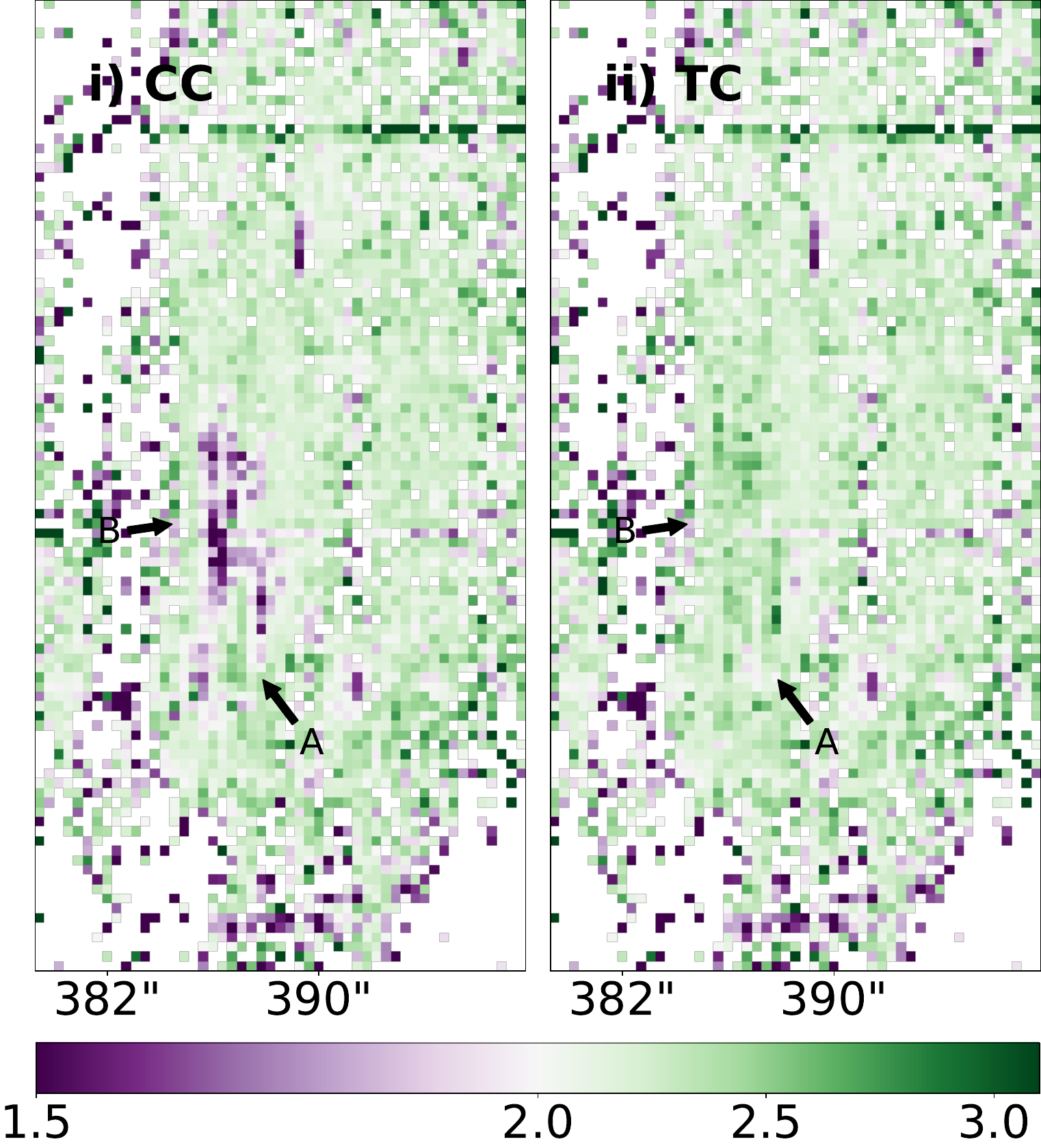}
    \end{subfigure}
    
    \begin{subfigure}[c]{0.25\linewidth}
        \centering
        \caption{Jet 2 Total Intensity}
        \label{rad jet 2}
        \includegraphics[width=.85\linewidth]{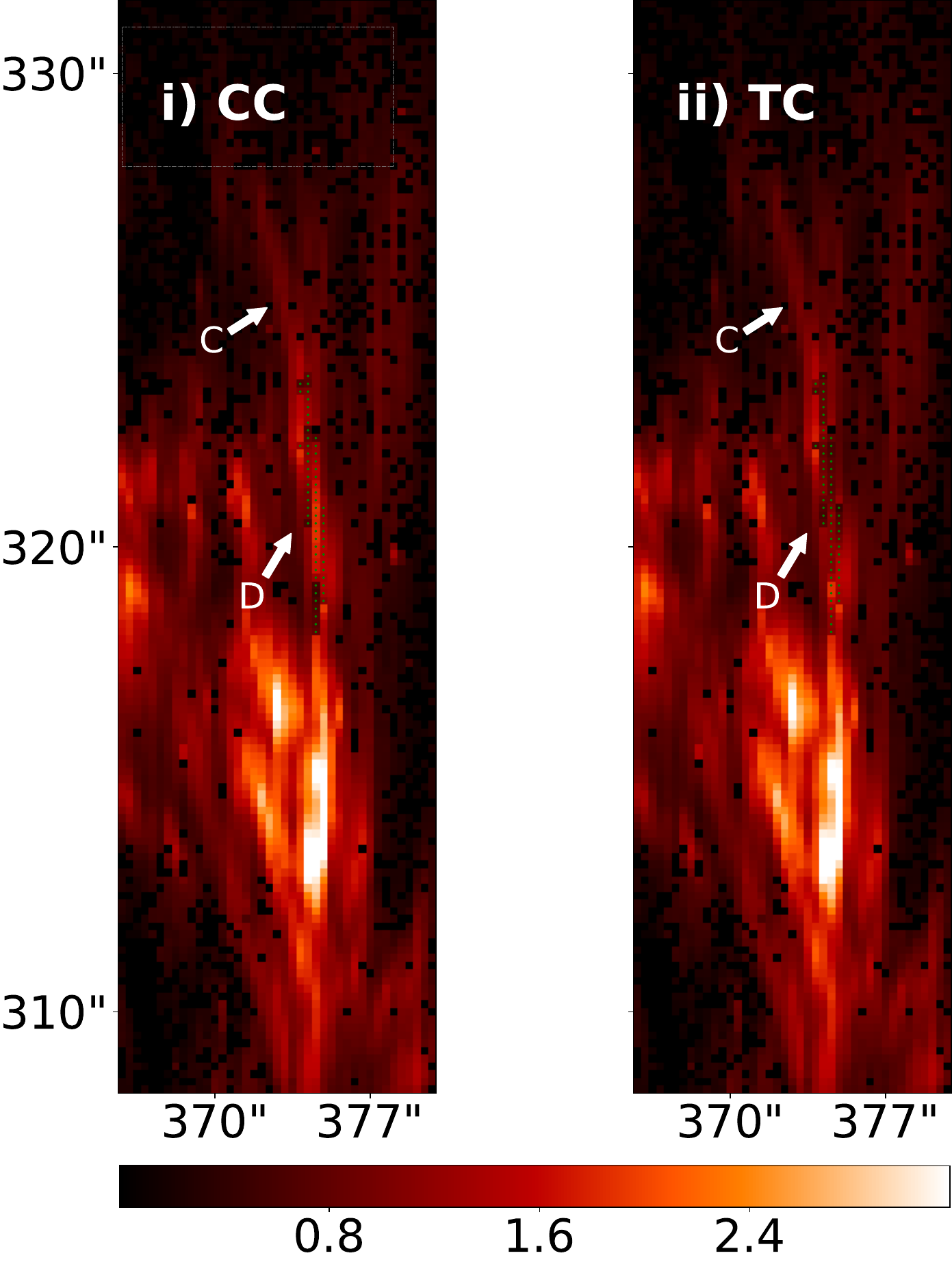}
    \end{subfigure}
    \begin{subfigure}[c]{0.22\linewidth}
        \centering
        \caption{Jet 2 Doppler Velocity}
        \label{doppler jet 2}
          \includegraphics[width=.85\linewidth]{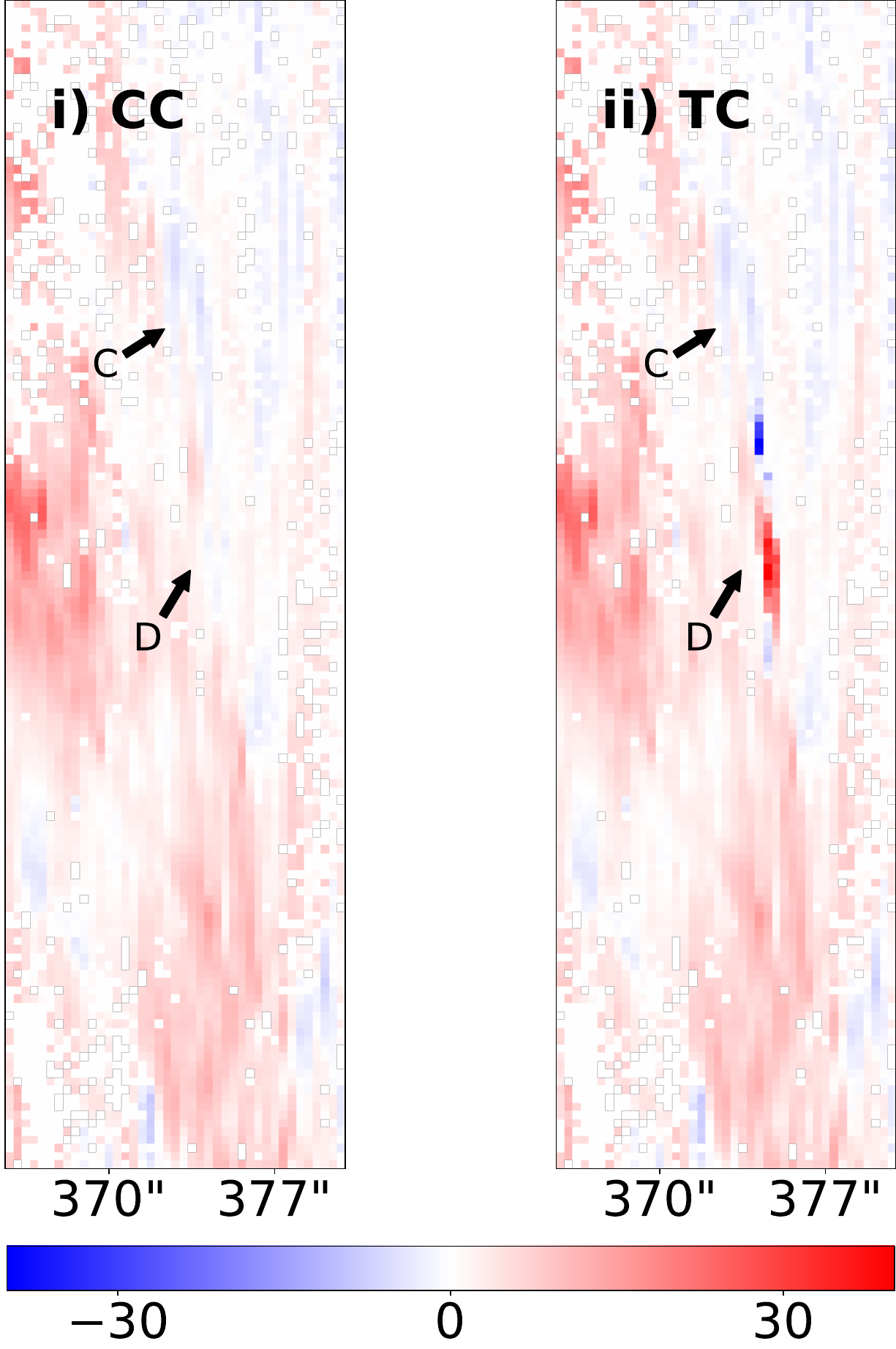}
    \end{subfigure}
    \begin{subfigure}[c]{0.22\linewidth}
        \centering
        \caption{Jet 2 Nonthermal Velocity}
        \label{nonthermal jet 2}
                \includegraphics[width=.85\linewidth]{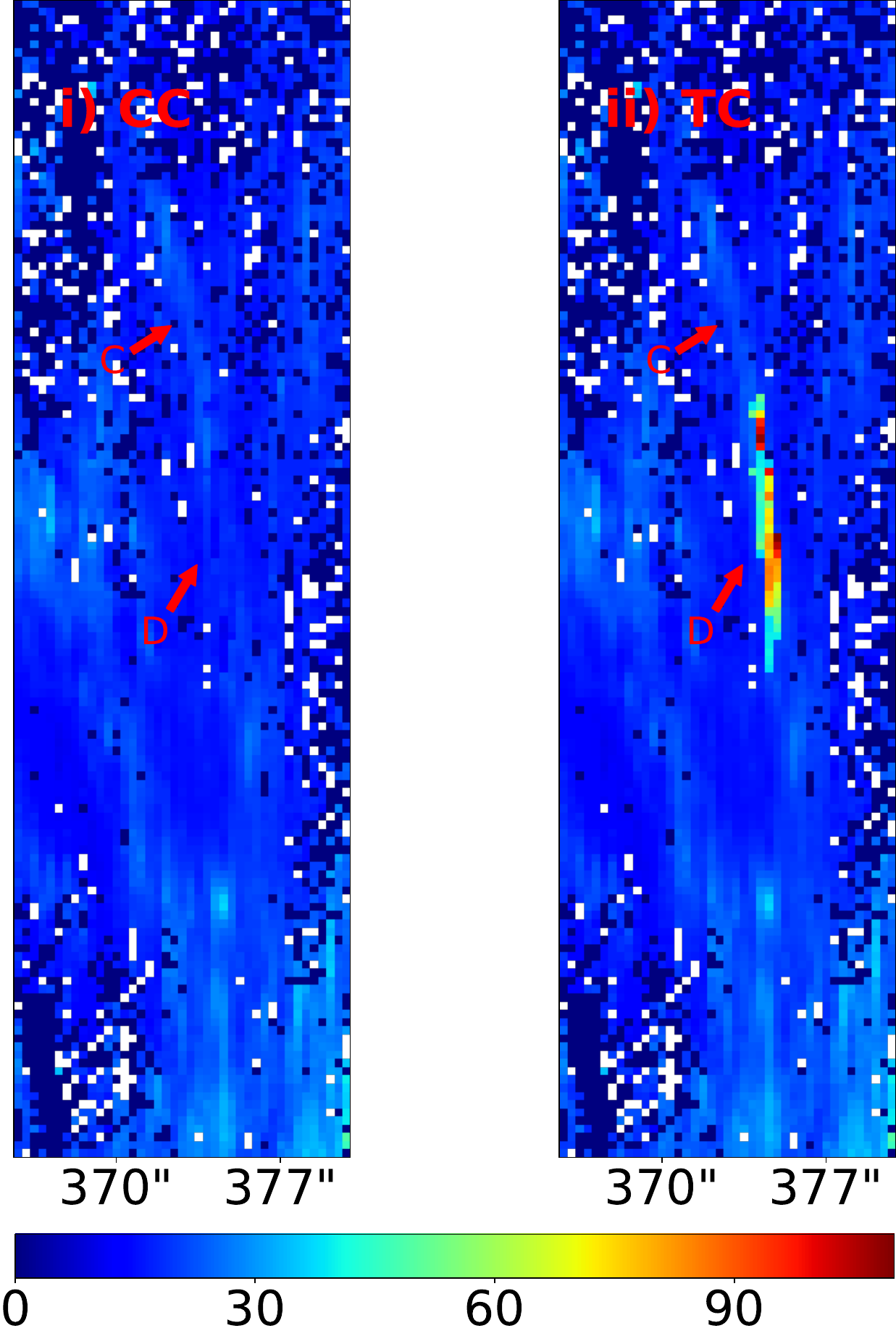}
    \end{subfigure}
    \begin{subfigure}[c]{0.225\linewidth}
        \centering
        \caption{Jet 2 Ratio}
        \label{ratio jet 2}
\includegraphics[width=.85\linewidth]{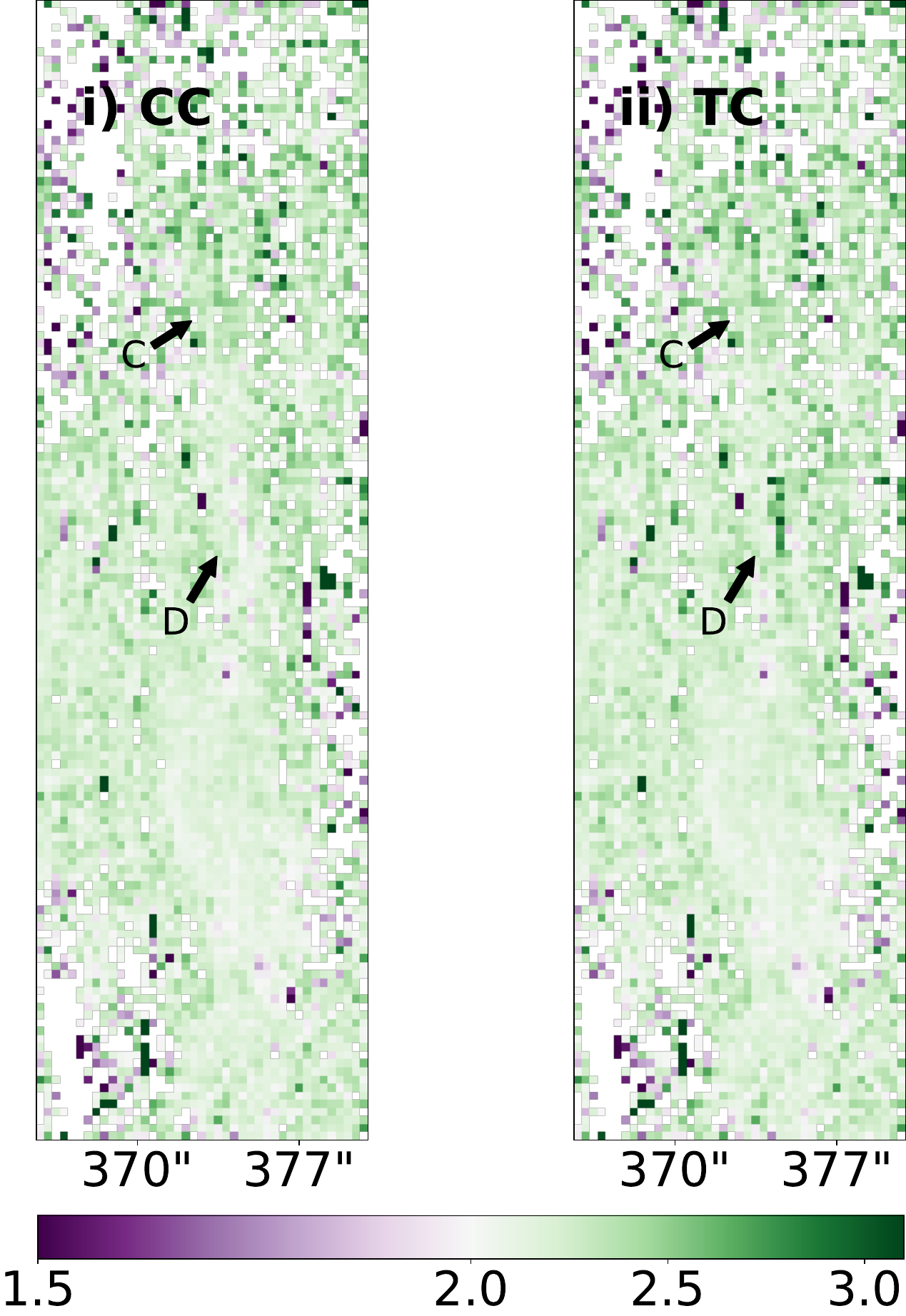}
    \end{subfigure}
    
    \caption{Maps of Doppler velocity in km/s, nonthermal velocity in km/s, total linear intensity in 10$^3$ $\times$ erg s$^{-1}$ sr$^{-1}$ cm$^{-2}$, and the 1393.755~\AA/1402.770~\AA\ ratio of jet 1 (top 4 panels) and jet 2 (bottom 4 panels). The values are derived from the single Gaussian fit of the Si \textsc{IV} 1393.755~\AA\ spectral line, except the locations where profiles are best described by a bimodal Gaussian fit and shown with the green points in panels (a) and (e). In the left panel of each subfigure, the values of the green-marked pixels are those of the core component (marked CC) of the bimodal Gaussian fit, while in the right panel these values indicate those of the tail component (marked TC). The gray boxes in panels (a) and (e) show the regions that we define as background in \autoref{intensity}. Lastly, the arrows show the position of the jet 1 base (arrow A), the loop associated with jet 1 (B), the jet 2 spire (C), and jet 2 base (D).}
    
    \label{raster}
\end{figure*}

In the following subsection, we present an analysis of the images in \autoref{ratio jet 1} and \autoref{ratio jet 2}, which show the intensity ratios calculated for jets 1 and 2, respectively. The ratios are calculated as follows: The total (integrated) intensity of each spectral profile in counts (DN) is converted to flux in
physical units (erg s$^{-1}$ sr$^{-1}$ cm$^{-2}$). The intensity ratio is calculated for the single Gaussian as:

\begin{equation}
   ratio = \frac{I_{1393.755\text{\AA}}}{I_{1402.770\text{\AA}}}
\end{equation}

For each bimodal Gaussian component, we use the following quantities: 

\begin{equation}
r_{CC} = \frac{I_{CC,1393.755\text{\AA}}}{I_{CC,1402.770\text{\AA}}}
\end{equation}

\begin{equation}
r_{TC} = \frac{I_{TC,1393.755\text{\AA}}}{I_{TC,1402.770\text{\AA}}}
\end{equation}

for the core components and  the tail components. The application of constraints in the fitting process, as explained in \autoref{fitting}, ensures that no bias is inserted into the calculation of the ratios.

We observe locations where r$_{CC}$ and r$_{TC}$  deviate from 2, indicating that opacity or resonance scattering effects become important. For jet 1 (\autoref{ratio jet 1}), we notice an area at $(386\arcsec, 274\arcsec )$, located at the loop adjacent to the jet, where r$_{CC}$ < 2. This indicates plasma with important opacity in the loop structure. On the other hand, values r$_{CC}$ > 2 and r$_{TC}$ > 2 at (X,Y)=(386\arcsec, 272\arcsec) are observed and can be attributed to the bright sources of the tail component, causing resonance scattering, and located at the base of jet~1, along with areas in the (X,Y)=(386\arcsec, 271\arcsec) (\autoref{rad jet 1} panel b). Moreover, $r_{TC}\simeq\ 2 $ in these regions of high intensity in \autoref{rad jet 1}, probably because the thermal collision component dominates the emission there. Finally, r$_{TC}$ > 2 and r$_{CC}$ < 2 for large areas, ranging from $(387\arcsec, 271\arcsec )$ to $(387\arcsec, 278\arcsec )$.
For jet 2 (\autoref{ratio jet 2}), locations with  r$_{TC}$ > 2 are present at the jet base and the jet spire. In most cases, $r_{CC}\simeq\ 2 $.

In general,  the regions where r$_{TC}$ > 2 seem to those where the tail component presents high nonthermal velocities. Such examples are the base of jet~1 and around the base of jet~2 (375\arcsec, 320\arcsec).
A more thorough quantitative study with more jet examples is necessary to check if our results may be generalized.

\subsection{Temperature}

We use the EM-loci method
\citep{jordan_1987, Patsourakos_2007} to estimate the temperature of the jets:

\begin{equation}
    EM=\frac{I_{obs}}{Resp(T)}, 
\end{equation}

where the observed intensity I$_{obs}$ at each wavelength range corresponds to a number of spectral lines along with a continuum and Resp(T) represents the instrument response function for that given wavelength, which is obtained using the Interactive Data Language (IDL) procedure  aia$\_$get$\_$response.pro.
In order to minimize uncertainties due to differences in element abundances, we use AIA channels observing primarily iron ions. We also use coronal abundances \citep{feldman_abundances} following the CHIANTI 10.0.1 library.

\begin{figure}
    \centering
   \resizebox{\hsize}{!}{\includegraphics{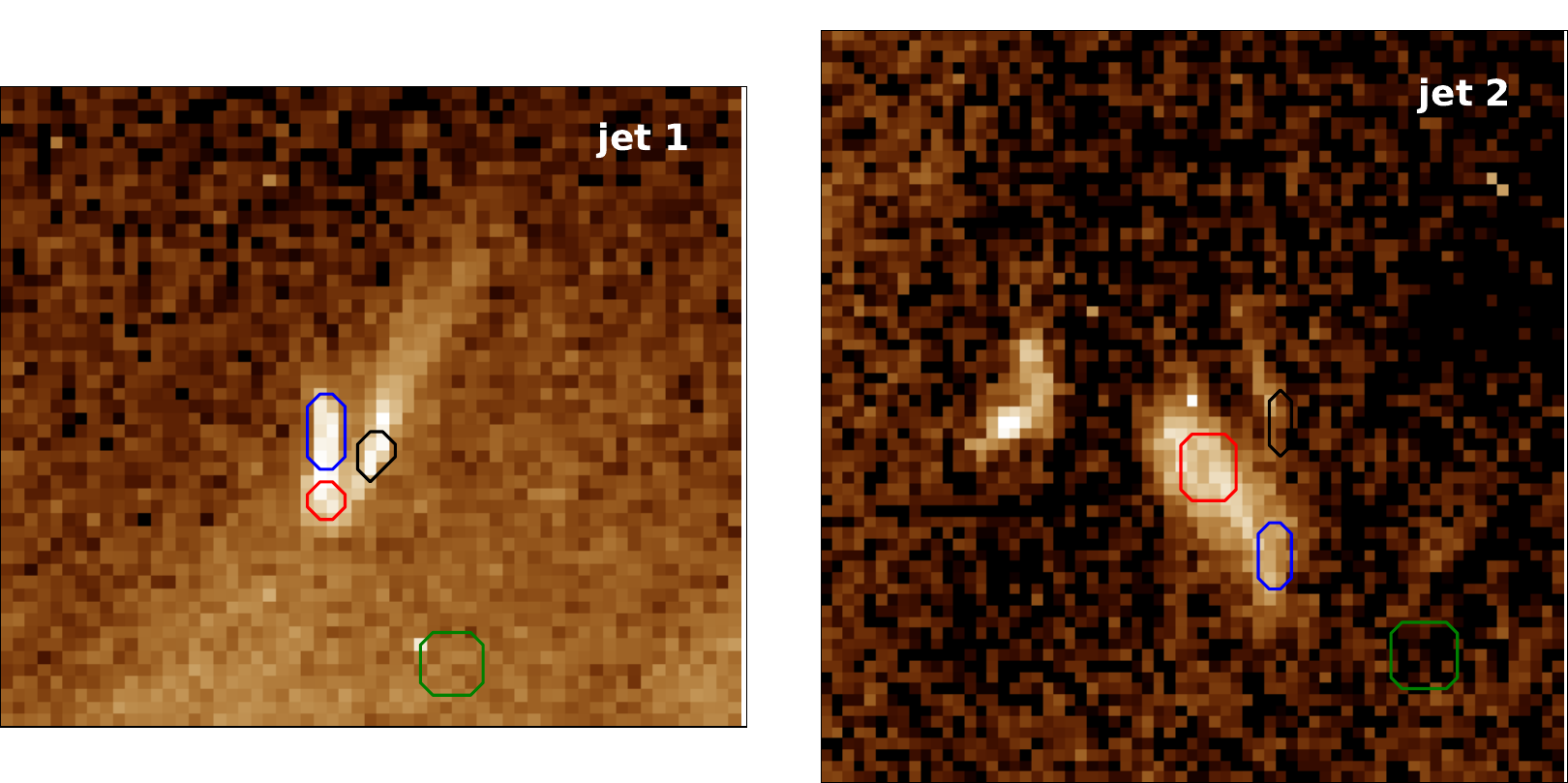}}
    \caption{AIA 193~\AA\ images of the two jets. The contours show the areas for which the EM-loci curves are calculated and shown in \autoref{em loci}. For jet 1, black = spire, red = loop A, blue = loop B, green = plume (foreground). For jet 2, black = spire, red = coronal bright point A, black = coronal bright point B, and green = background.}
    \label{em map}
\end{figure}

\begin{figure}
\centering
\resizebox{\hsize}{!}{
\includegraphics{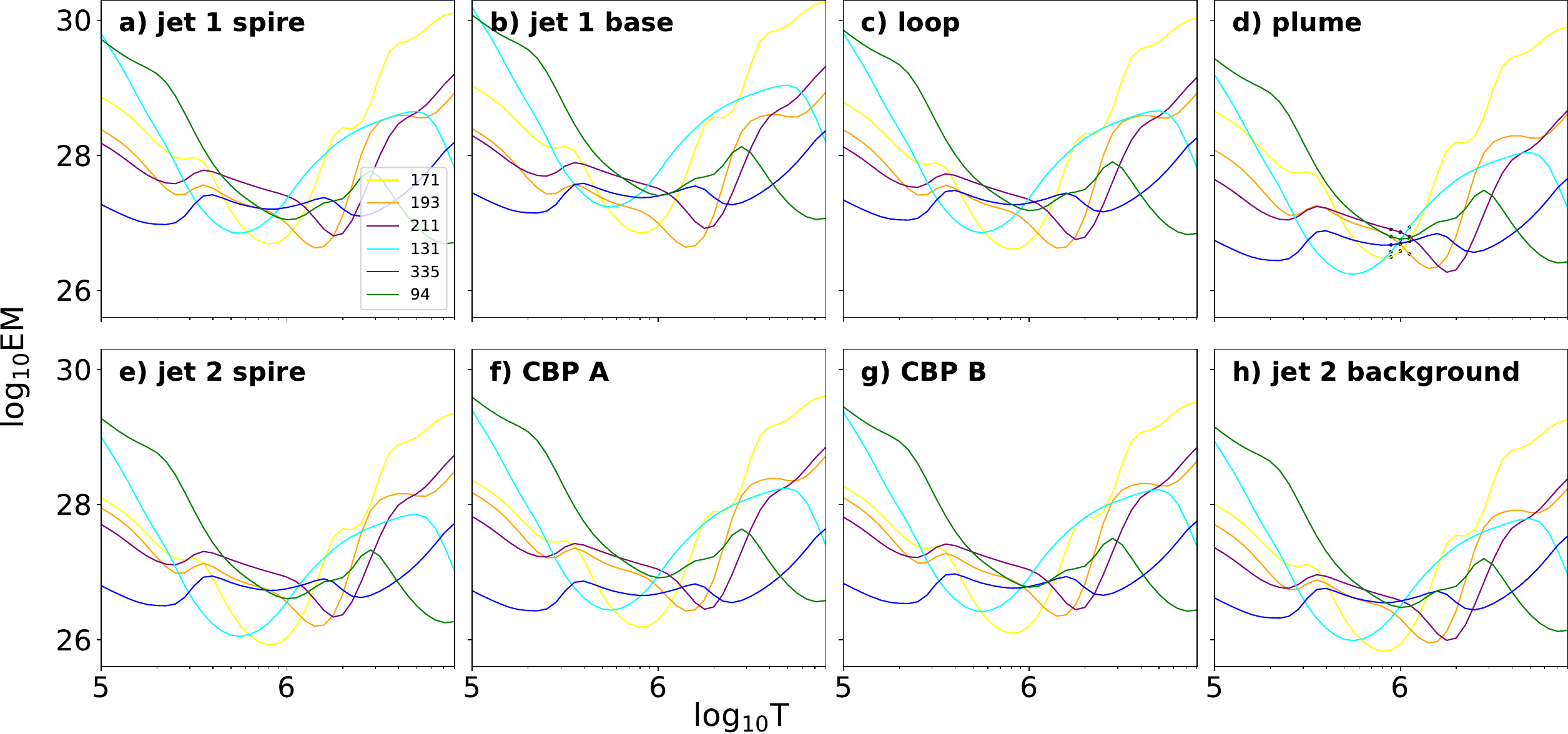}}
\caption{EM-loci curves of the two areas. EM and temperature (T) are measured in cm$^3$ and K, respectively. The top row corresponds to jet 1, and shows the jet spire (panel a), the jet base (panel b), a region on the loop (panel c), and the plume emission (panel d) at 17:27~UT,  . The bottom row corresponds to jet 2, and shows the spire (panel e), two regions on the coronal bright point (panels f and g), and the background (panel h) at 17:08~UT. The black dots in panel (d) correspond to the points around each minimum where each EM curve is in closest proximity with the others. }
 \label{em loci}
\end{figure}

We selected the intensities of these structures in all six coronal AIA wavelengths used in this study. Jet~1 is not visible in  94~\AA,\  while jet~2 is absent from
94~\AA\ and 335~\AA, and thus the sampled intensities in these wavelengths represent the upper limit of the structure emission.
We calculated the EM-loci curves for four subregions of each area. For jet 1, these regions are the plume foreground, the loop, the jet base, and the jet spire, while for jet 2, these are the background, the location "A" on the coronal bright point, the location "B" on the coronal bright point and the jet spire(see \autoref{em map}). The EM-loci curves are shown in \autoref{em loci}. We observe that the EM-loci curves of jet 1 (panels a to c in \autoref{em loci}) exhibit higher values than those of jet 2 (panels e to g in \autoref{em loci}), confirming that jet 1 is brighter than jet 2. For example, jet 1 shows higher EM values by a factor of 7 for 171~\AA,
by a factor of 8 for 131~\AA,\ and by a factor of 3.5 for 94~\AA.

In an ideal situation, for a purely isothermal plasma, the EM curves would have a single point of intersection corresponding to the plasma temperature. In \autoref{em loci}, the only case indicating the presence of isothermal plasma is the plume (panel d), as the EM curves calculated from the plume have similar values for T$\sim$10$^6$K. In the other panels, for the features composing the two jets and the background, the plasma appears to be multithermal as there are no temperatures where all EM curves present similar values. 

In order to further investigate whether each region is consistent with isothermal or multithermal plasma, we calculate the average logarithmic difference $\Delta (log_{10}EM)$: 

\begin{equation}\label{delta_em_eq}
    \Delta (log_{10}EM) = \frac{1}{N} \sum_{i=1}^N \left ( \sum_{j=i+  1}^N |log_{10}EM_i - log_{10}EM_j | \right )
,\end{equation}

where N is the number of AIA channels and i=1,2,3,4,5,6 correspond to wavelengths 94~\AA, 131~\AA, 171~\AA, 193~\AA, 211~\AA, and 335~\AA. 

Ideally, $\Delta (log_{10} EM)$ would be equal to zero for a specific temperature for an isothermal plasma. In a real observation, this result would be affected by the presence of errors.
Assuming that all channels exhibit a net uncertainty of 35$\%$ \citep{Guennou_2012}, we can estimate the uncertainty due to calibration errors, photometric errors, and so on as $\Delta (log_{10}EM)_{uncert}=log_{10} \left( \frac{1.35}{0.65}\right)\approx 0.3$, which gives the maximum possible uncertainty \citep{Patsourakos_2007}.

Our results are consistent with isothermal plasma if $\Delta (log_{10}EM)$ < $\Delta (log_{10}EM)_{uncert}$. As mentioned above, the plume (jet 1 foreground, panel d) is the only region consistent with isothermal plasma with a temperature of  T = 10$^6$ $\pm$ 10$^5$K. The temperature error is calculated as the standard deviation of the temperatures corresponding to the points around the minimum where each EM curve is in closest proximity with all others \citep{Gupta_2015}. These points are shown with black dots in \autoref{em loci} panel d).
On the other hand, multithermal distributions are possible for all other regions. 

In order to estimate the range between a low temperature value T$_1$ and a high value T$_2$, we calculated the $\Delta (log_{10}EM)$ curves for synthetic rectangular differential emission measure (DEM) distributions (denoted $\Delta(log_{10}EM)_{DEM}$ hereafter), following \cite{Patsourakos_2007} . We take multiple combinations of temperatures T$_1$ and T$_2$, ensuring that the lowest temperature T$_1$ is not higher than 10$^{5}$K, as the response of AIA is very low below this temperature \citep{Viall_2020}. For each region, we select the combination of T$_1$ and T$_2$ that results in the lowest euclidean distance between the $\Delta (log_{10}EM)_{DEM}$ curve and the observed $\Delta (log_{10}EM)$. The best combinations of T$_1$ and T$_2$ for each region are presented in the following paragraph.

For jet~1, T$_1$ is equal to $10^5$~K and T$_2$ is equal to $10^{6.2}$~K for the spire, the base, and the loop. We believe that jet 1 is bright in 335~\AA\, because  this channel is sensitive to temperatures of log$_{10}$T=5.3 (see \autoref{aia_response_function}), which is included in the T$_1$ to T$_2$ range for this jet.  
Jet~2 has higher temperatures with T$_1$=10$^{5.7}$K at its spire and 10$^{5.4}$K for the coronal bright point, while  
temperature T$_2$ is $10^{6.4}$~K for both regions. Jet~2 seems to be slightly hotter than the background compared to jet 1.
Jet~2 is not visible in channels 94~\AA\ and 335~\AA\ as it is fainter rather than cooler compared to jet~1 (see \autoref{observation}).

\begin{figure}
\centering
\resizebox{\hsize}{!}{
\includegraphics[width=.99\linewidth]{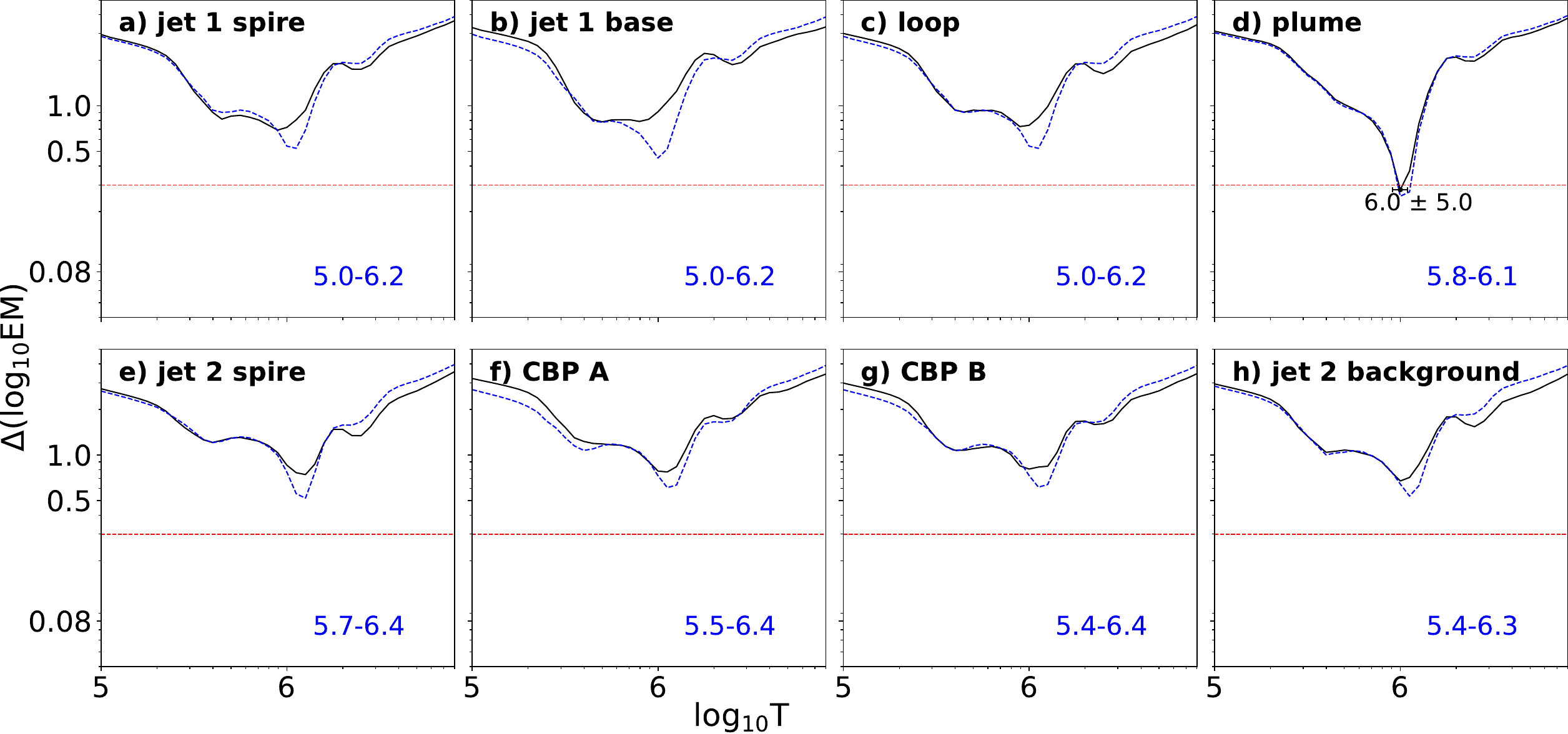}}

   \caption{The $\Delta log_{10}$EM curves for the two areas. Temperature is measured in K.
   The top row corresponds to jet 1, showing results for the jet spire (panel a), the jet base (panel b), a region on the loop (panel c), and the plume emission (panel d) at 17:27~UT. The bottom row corresponds to jet 2, showing the spire (panel e), two regions on the coronal bright point (A and B in panels f and g, respectively), and the background (panel h) at 17:08~UT. The red dashed horizontal line at 0.3 represents the net uncertainty due to calibration errors, photometric errors, and so on (see text for more details). In panel d, we annotate the temperature logarithm corresponding to the minima $\Delta$EM, as well as the logarithm  of the temperature error. Here, 6.0 $\pm$ 5.0 is a representation denoting 1.0 $\times$ 10$^{6.0}$ $\pm$ 1.0 $\times$ 10$^{5.0}$ K. The blue dashed lines correspond to the synthetic $\Delta \log_{10}$EM curves derived from the simple rectangular DEM distributions. The logarithmic temperature range of the DEM distribution for each region is shown in blue in the bottom-right corner.}
\label{delta em}
\end{figure}

\autoref{delta em} shows the observed $\Delta (log_{10}EM)$ curves for each region (black solid lines) and the $\Delta (log_{10}EM)_{DEM}$ curves for the synthetic rectangular DEM distributions, which in most cases successfully represent the observed  $\Delta$ (log$_{10}$ EM ) curves. The logarithmic temperature range of the DEM distribution for each region is shown in blue in the bottom-right corner. For example, in panel (a), 5.05-6.25 denotes a range of T$_1$=10$^{5.05}$ K to T$_2$=10$^{6.25}$ K. Finally, the red horizontal dashed line represents $\Delta (log_{10}EM)_{uncert}$.
We observe that the spires of the two jets exhibit similar temperature distributions.

\subsection{Electron density}

We calculated the electron density with the use of the O~\textsc{IV} 1399.77~\AA/1401.16~\AA\ line ratio, which can be found in the 1403~\AA\ IRIS spectral window. Assuming a near-isothermal plasma in ionization equilibrium, these intercombination lines are suitable for density diagnostics \citep{dudik_density}. 
In order to reduce noise and improve the fitting, we smoothed each profile with the Savitzky–Golay filter \citep{Savitzky1964}, which calculates a polynomial fit of each window based on a polynomial degree equal to 3 and window size equal to 11. The filter is provided by the scipy Python library \citep{2020SciPy-NMeth}. Subsequently, we fitted each spectral profile of the two lines separately with a single Gaussian curve, we calculated the integrated intensity, and performed radiometric calibration. The integrated line ratio is compared to the theoretical ratio retrieved from the CHIANTI 10.0.1 atomic database \citep{Del_Zanna_2021}.

\begin{figure}[hbt!]
    \centering
   \begin{subfigure}[c]{0.99\linewidth}
    \centering
        \caption{jet 1}
        \label{eletron density jet 1}
\includegraphics[width=\linewidth]{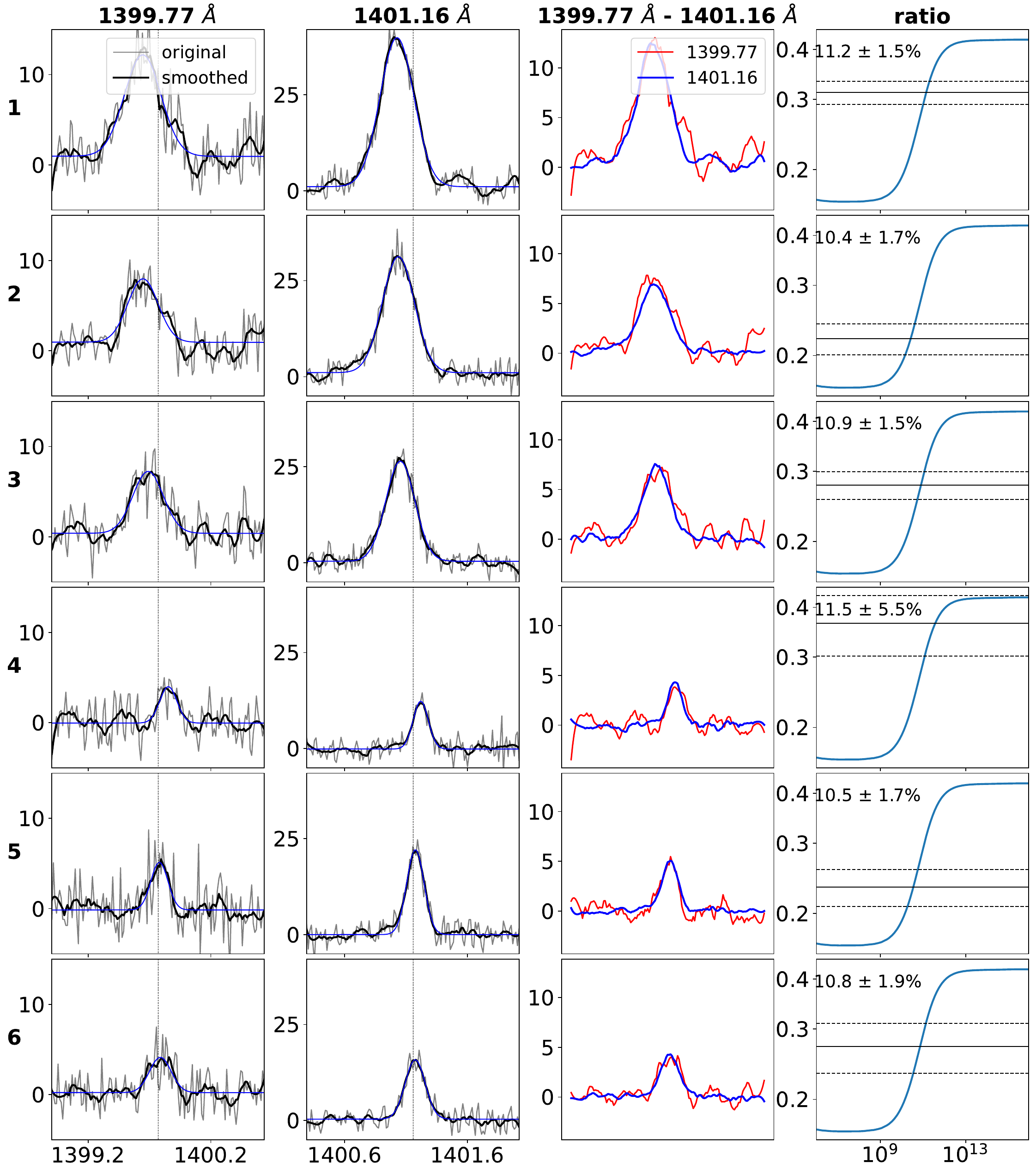}
   \end{subfigure}
   
    \begin{subfigure}[c]{0.99\linewidth}
    \centering
     \caption{jet 2}
     \label{eletron density jet 2}
    \includegraphics[width=\linewidth]{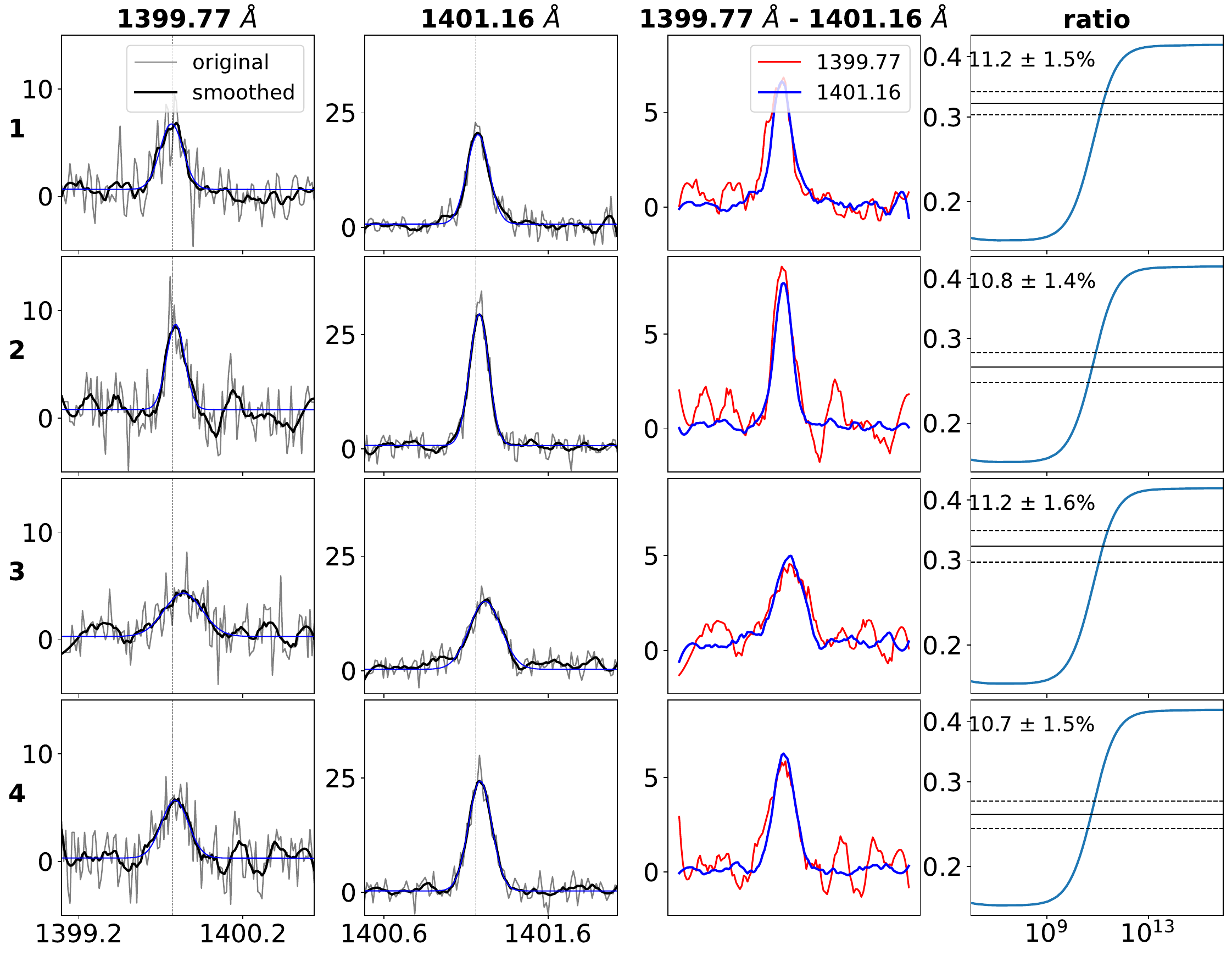}

          \end{subfigure}
   \caption{Electron density calculation for jet 1 and jet 2. Here, we show plots of O~\textsc{IV} 1399.77~\AA\ (first column) and O~\textsc{IV} 1401.16~\AA\ (second column), comparative plots of  O~\textsc{IV} 1399.77~\AA\ and  O~\textsc{IV} 1401.16~\AA\ multiplied by the ratio (third column), and the theoretical ratio--electron density plot by CHIANTI 10.0.1. (forth column), where the solid horizontal line is the calculated ratio and the dashed ones are the upper and lower values. Intensity is measured in erg s$^{-1}$ sr$^{-1}$ cm$^{-2}$ \AA $^{-1}$ and density in cm$^{-3}$.}
    \label{eletron density}
   \end{figure}

\begin{figure}
\centering
  \resizebox{\hsize}{!}{
\includegraphics[width=.99\linewidth]{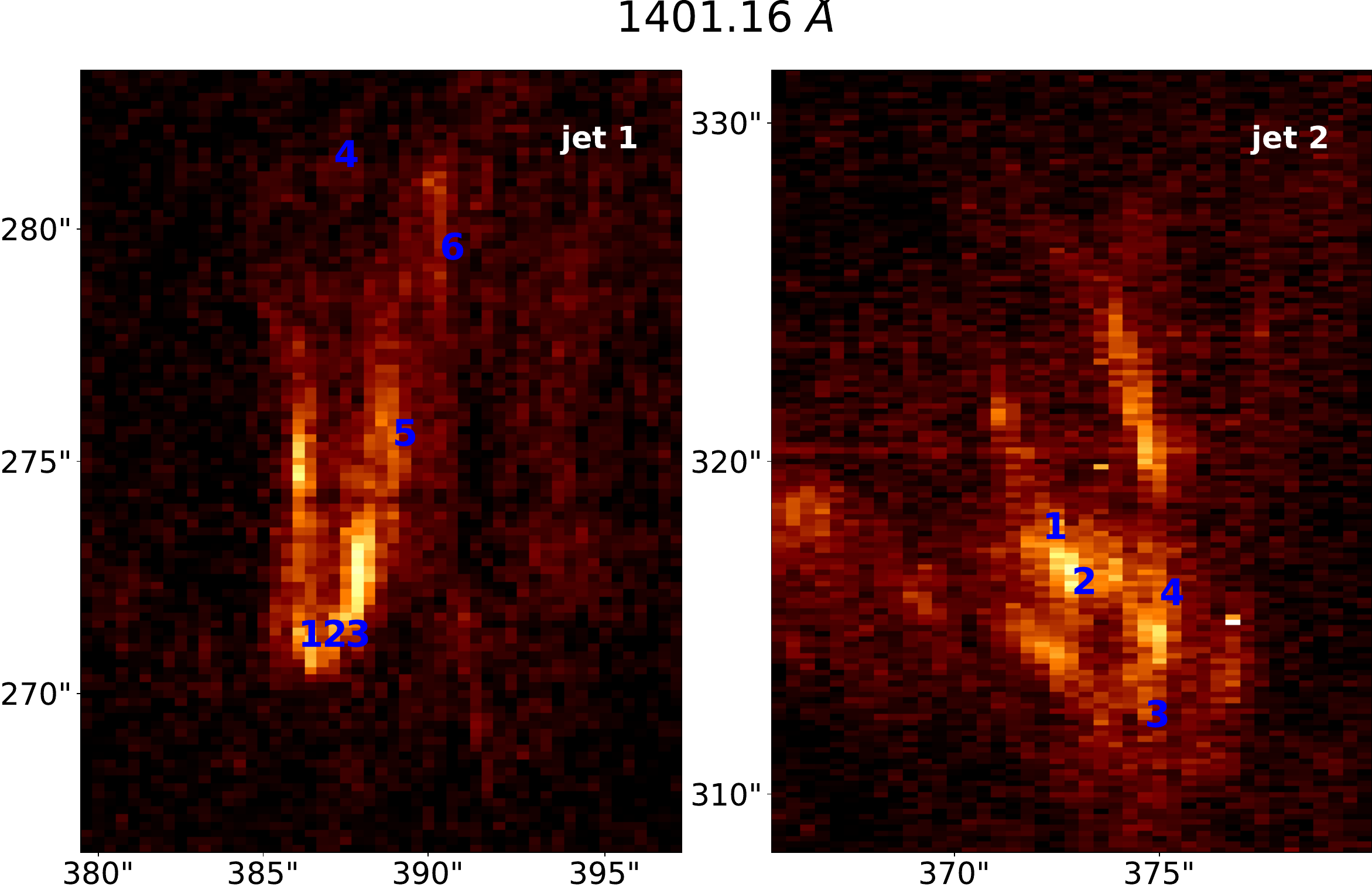}}
   
   \caption{
    Total intensity map measured in $erg s^{-1} sr^{-1} cm^{-2}$ of O~\textsc{IV} 1401.16~\AA\ of jet  1 (left panel) and jet 2 (right panel), where the exact position of each spectral profile is indicated with a number corresponding to a row in \autoref{eletron density jet 1} and \autoref{eletron density jet 2}.}
\label{electron density map}
\end{figure}

\autoref{eletron density} shows the spectral profiles that were used for the electron density calculation for each jet. 
The figure shows the spectral profiles of O~\textsc{IV} 1399.77~\AA\ (first column) and O~\textsc{IV} 1401.16\AA\ (second column),  comparative plots of  O~\textsc{IV} 1399.77~\AA\ and  O~\textsc{IV} 1401.16~\AA\ multiplied with the ratio (third column), and a theoretical ratio--electron density plot by CHIANTI 10.0.1. (forth column). The solid horizontal line is the calculated ratio and the dashed ones are the upper and lower values, taking into account the propagation of error. Intensity is measured in erg s$^{-1}$ sr$^{-1}$ cm$^{-2}$ \AA$^{-1}$ and density in cm$^{-3}$.  
\autoref{electron density map} shows the total intensity map of O~\textsc{IV} 1401.16~\AA\ for each jet, where the exact position of each spectral profile is indicated with a number corresponding to a row of panels of \autoref{eletron density}. 
In the third column of panels of \autoref{eletron density}, one can see that the profiles of the O \textsc{IV} 1399.77~\AA\  line multiplied by the ratio correspond well  to the those of O \textsc{IV} 1401.16~\AA,\ which is qualitative evidence of the influence of the noise in the profiles \citep[e.g.,][]{Young_2015}. 

For jet 1, we find densities of from 10$^{10.2}$ to 10$^{11.3}$ cm$^{-3}$ on the jet-loop base (profiles 1,2,3) and 10$^{10}$ cm$^{-3}$ at the jet spire (profile 5). The plasma in the surrounding region appears to be denser (profile 4), reaching 10$^{11}$  cm$^{-3}$; however, the error on this calculation is significant, as the upper value does not fall into the linear segment of the theoretical density--ratio curve. 

For jet 2, we measure a density of  10$^{10.6}$-10$^{11.4}$ cm$^{-3}$ at the coronal bright point. Unfortunately, no measurements at the jet spire were possible, because no reliable Gaussian fittings were derived from the O \textsc{IV} 1399.77~\AA\ profiles at this region.

\section{Discussion and conclusions} \label{discussion}

\begin{table}
\centering
\caption{Parameter values calculated for the two studied jets.}

\begin{threeparttable}
\resizebox{\hsize}{!}{%
\begin{tabular}{c c c c c}
\hline\hline 
 \textbf{Parameter} & \multicolumn{2}{c}{\textbf{Jet 1}} & \multicolumn{2}{c}{\textbf{Jet 2}} \\ 
 \hline
\multicolumn{1}{c}{\textbf{plane-of-sky length \tnote{a}}} & \multicolumn{2}{c}{10.5} & \multicolumn{2}{c}{9} \\ 
\multicolumn{1}{c}{\textbf{plane-of-sky width \tnote{a}}} & \multicolumn{2}{c}{2} & \multicolumn{2}{c}{1.5} \\ 
\multicolumn{1}{c}{\textbf{duration \tnote{b}}} & \multicolumn{2}{c}{4-6} & \multicolumn{2}{c}{3-4} \\ 
\multicolumn{1}{c}{\textbf{plane-of-sky speed \tnote{c}}} & \multicolumn{2}{c}{100} & \multicolumn{2}{c}{60} \\ 
\multicolumn{1}{c}{\textbf{TR spire density \tnote{d}}} & \multicolumn{2}{c}{10$^{10}$} & \multicolumn{2}{c}{-} \\ 
\multicolumn{1}{c}{\textbf{TR base density \tnote{d}}} & \multicolumn{2}{c}{10$^{11}$} & \multicolumn{2}{c}{10$^{11}$} \\ 
\multicolumn{1}{c}{\textbf{spire temperature \tnote{e}}} & \multicolumn{2}{c}{$10^{5.0}$ - $10^{6.2}$} & \multicolumn{2}{c}{$10^{5.7}$ - $10^{6.4}$} \\ 
\multicolumn{1}{c}{\textbf{base temperature \tnote{e}}} & \multicolumn{2}{c}{$10^{5.0}$ - $10^{6.2}$} & \multicolumn{2}{c}{\textit{$10^{5.4}$ - $10^{6.4}$ (CPB)}} \\ 
\multicolumn{1}{c}{\textbf{background temperature \tnote{e}}} & \multicolumn{2}{c}{\textit{$10^{6}$ (plume)}} & \multicolumn{2}{c}{$10^{5.4}$ - $10^{6.3}$} \\  

 \hline
\multicolumn{1}{c}{} & \bf{CC} & \bf{TC} & \bf{CC} & \bf{TC} \\ 

\multicolumn{1}{c}{\textbf{mean redshift DV \tnote{c}}} & 8 & 16 & \textbf{5} & \textbf{11} \\ 
\multicolumn{1}{c}{\textbf{max redshift DV \tnote{c}}} & 23 & 67 & 13 & 42 \\ 
\multicolumn{1}{c}{\textbf{mean blueshift DV \tnote{c}}} & -5 & -21 & -1 & -8 \\ 
\multicolumn{1}{c}{\textbf{max blueshift DV \tnote{c}}} & -19 & -83 & -3 & -42 \\ 
\multicolumn{1}{c}{\textbf{mean NTV \tnote{c}}} & 22 & 55 & 15 & 51 \\ 
\multicolumn{1}{c}{\textbf{max NTV \tnote{c}}} & 42 & 134 & 21 & 110 \\ 
\multicolumn{1}{c}{\textbf{mean total Intensity \tnote{f}}} & 1500 & 2800 & 1100 & 1200 \\ 
\multicolumn{1}{c}{\textbf{max total Intensity \tnote{f}}} & 6500 & 8200 & 2000 & 3300 \\ \hline
\end{tabular}}
\begin{tablenotes}[para,flushleft] 
 \parbox[t]{0.95\linewidth}{\textbf{Notes.} DV and NTV denote Doppler velocity and nonthermal velocity, respectively. CBP denotes coronal bright point. The length, width, duration, and plane-of-sky speed are calculated from 193~\AA\ AIA images. The Doppler and nonthermal velocities, as well as the intensity values, are calculated from the 1393.775~\AA\ Si \textsc{IV} spectral line. The core and tail component values from the bimodal profiles are described as CC and TC, respectively.

\item[a] Mm, \quad \item[b] minutes, \quad \item[c] for the 1393.775~\AA\ line in km/s, \quad \item[d] cm$^{-3}$, \quad \item[e] at the corona, in K, \\ \quad \item[f] for the 1393.775~\AA\ line in erg s$^{-1}$ sr$^{-1}$ cm$^{-2}$}
\end{tablenotes}
\end{threeparttable}

\label{jet_parameters}
\end{table}

We studied two jets originating from an on-disk CH, combining AIA images and IRIS spectroscopic data. We performed several measurements of size, duration, plasma velocity, temperature, and electron density for both jets; our results are summarized in \autoref{jet_parameters} and are discussed below.

Both jets are associated with a bright structure at their base: a small loop  for jet 1
and a coronal bright point for jet 2. Using AIA 193~\AA\ images, we conclude that the plane-of-sky length of jet 1 is roughly twice the plane-of-sky \textbf{length} of the small loop at its base,
while jet 2 is significantly smaller than the coronal bright point. Jet 1 is brighter than jet 2, has almost twice its duration, and twice the magnitude of  plane-of-sky speed.  
Moreover, jet 1 is associated with the eruption of a small mini-filament.

The asymmetrical Si \textsc{iv} profiles in both jets can be attributed, in many cases, to the overlap of background
emission and high-speed flows \citep{Gorman2022}. For both jets, the core-component features of the bimodal Gaussian profiles are, in most cases, comparable to  single-Gaussian profiles found outside the jets, which we use to represent the background. On the other hand, the tail component features are associated with the most dynamic parts of the jets and their connected structures, such as a small loop for jet 1 and the coronal bright point. 

The Doppler shifts measured with the tail spectral component features of Si \textsc{IV}  1393.755~\AA\ and 1402.770~\AA\ lines are much higher for jet 1. Combined with the plane-of-sky speed measured with the AIA 193~\AA\ images, we conclude that jet 1 is faster than jet 2.
Interestingly, the nonthermal velocities measured with the tail component spectral features of Si \textsc{IV}  1393.755~\AA\ lines reach similar maximum and mean values
at the bases of the two jets, as well as at one of the loop footpoints of jet 1. Assuming that the nonthermal velocity is related to turbulence or Alfv\'en waves, this can be evidence of similar energy rates for the two jet eruptions \citep{Chae_1998}; although they differ in other physical characteristics.
However, for jet 1, the presence of a tail component is found in 1.5 times more individual profiles (see \autoref{fitting})
and the peak intensities of the tail component of jet 1 are two to four times higher than the corresponding peak intensities of jet 2 (see \autoref{profile_examples}).
This indicates that a much larger plasma volume is affected by turbulent energy (or wave energy) in jet 1 than in jet 2, despite the fact that the corresponding energy flux  for both jets is comparable.

The line-ratio analysis can provide information on  the plasma emission.
An interesting finding is that r$_{TC}$>2 at the base of the jets.
Moreover, the r$_{TC}$ > 2 locations are associated with high nonthermal velocity values ($v_{nth}$ > 70 km/s) of the tail component. The locations with r$_{TC}$ > 2 indicate the effects of resonant scattering and can be explained by 
the presence of a bright radiation source, such as at the locations at the base of jet~1. 
Resonant scattering can also be caused by reduced thermal emission due to relatively low electron density or a plasma temperature associated with a low Si \textsc{IV} ion population. 
This may be the case at the r$_{TC}$ > 2 locations of jet~2, where the intensity of the structures surrounding these locations ---which may represent the incident intensity at these locations--- is not as high as for jet 1. However, we note that a high ratio can also be caused by opacity combined with the specific geometry of the jet feature \cite{Keenan_2014}. These qualitative results regarding the jet bases must be confirmed with the analysis of more jet observations and detailed modeling of the emissions. Finally, let us mention that the IRIS spectral calibration  may cause systematic errors in the line ratio calculation \citep{Wuelser_2018}.

We measure TR electron densities of $10^{10}$ cm$^{-3}$ at the spire of jet 1,
while the base of the jet is denser, reaching 10$^{11}$ cm$^{-3}$, values close to those found by \cite{cai_2019}, who used the same line ratio at
an active region jet. The CPB from which jet 2 originates has an average density of 10$^{11}$ cm$^{-3}$, which is slightly higher than the
values  reported by \cite{Madjarska2019}. We conclude that the electron densities are similar 
at  the bases of both jets ---taking into account the associated errors. However, we have a small number of measurements and the aforementioned values are only indicative.

Employing the EM-loci method, we examined the temperature range of the studied structures.
The plume is very close to isothermality, with a temperature of 10$^6$K, in agreement with other observations \citep{Poletto2015}. Calculating simple rectangular DEM distributions, we also made a crude estimation of the temperature range for all other regions that are more consistent with multithermality, following \cite{Patsourakos_2007}.
Jet 1 exhibits a higher temperature range from 10$^5$ K up to 10$^{6.2}$ K at its spire, base, and loop. This result is in accord with the fact that we find a cooler component of the jet at 304~\AA\ and a hot component at the corona.
Similar results were obtained by \cite{Mulay_2017}, who, from a DEM analysis, identified cooler and hotter emission at the spire and footpoints of an active-region jet, respectively. The spire of jet 2 shows a temperature range of 10$^{5.7}$ K - 10$^{6.4}$ K, while the CPB ranges from 10$^{5.4}$ K to 10$^{6.4}$ K, similarly to the background. In conclusion, jet 2 appears to be hotter than jet 1, as the upper temperature given by the DEM analysis is higher than the corresponding temperature of jet 2; however jet 1 exhibits a broader temperature range.

Although we cannot deduce the exact mechanism responsible for the triggering of the jets, the HMI temporal evolution (\autoref{jet 1} panels 4-a to 4-f and \autoref{jet 2} panels 4-a to 4-d) provides evidence of slowly evolving photospheric magnetic structures of opposite polarity. We emphasize that any variation of the magnetic field strength is within the random noise of the instrument, as discussed in \autoref{observation}, and therefore we cannot make firm conclusions in regards to the magnetic flux evolution related to the jets given the small sizes of the structures. However, we do detect a weak polarity inversion line ($\pm$ 20 G) formed below the mini-filament associated with the eruption of jet 1, in accordance with the findings of similar studies \citep{Panesar_2018}.

In conclusion, we observe two jets that originate in two different areas of an on-disk coronal hole. The jet originating at the coronal bright point is slower, less bright, and smaller in AIA 193~\AA\ images compared to the jet that appears to be associated with a small mini-filament. Jet 2 appears to be hotter than jet 1, although jet 1 exhibits a broader temperature range. However, they have similar nonthermal velocities, reaching 110-130 km/s, and densities of $\sim$ 10$^{11}$ cm$^{-3}$ at their bases.

The precise role of these structures in the solar wind remains to be fully elucidated, and this will be the subject of our forthcoming research efforts. A more comprehensive study should include the analysis of the energy budget of the jet eruption, taking into account the magnetic energy accumulated and transferred via the magnetic field motions at the base of the jets. Furthermore, we aim to extend our study to a larger statistical sample.

\begin{acknowledgements} 

S. Patsourakos acknowledges support by the ERC Synergy Grant (GAN: 810218) ‘The Whole Sun’.
CHIANTI is a collaborative project involving George Mason University, the University of Michigan (USA), University of Cambridge (UK) and NASA Goddard Space Flight Center (USA). 

Costis Gontikakis acknowledges support by the research programme No 200/1021 of the Research Committee of the Academy of Athens.

We thank the referee for their insightful comments.

\end{acknowledgements}

\bibliographystyle{aa.bst} 
\bibliography{references.bib}

\end{document}